\documentclass[a4paper,11pt]{article}
\pdfoutput=1 

\usepackage{jcappub} 
\usepackage[T1]{fontenc} 
\usepackage{graphicx}
\usepackage{booktabs}
\usepackage{amsmath,amssymb,amsbsy,amstext, amsthm, amsfonts}
\usepackage{bm}
\usepackage{physics}
\usepackage{comment}
\usepackage{appendix}
\usepackage{xcolor}
\usepackage{soul}

\newcommand{\be}{\begin{eqnarray}}
\newcommand{\ee}{\end{eqnarray}}

\def\lsim{\mathrel{\raise.3ex\hbox{$<$\kern-.75em\lower1ex\hbox{$\sim$}}}}
\def\gsim{\mathrel{\raise.3ex\hbox{$>$\kern-.75em\lower1ex\hbox{$\sim$}}}}

\title{Supermassive Primordial Black Holes From Inflation}

\author[a,b,c]{Dan Hooper,}
\author[d]{Aurora Ireland,}
\author[a,b,c]{Gordan Krnjaic,}
\author[c]{\hspace{20mm}Albert Stebbins}

\affiliation[a]{University of Chicago, Kavli Institute for Cosmological Physics, Chicago, IL USA}
\affiliation[b]{University of Chicago, Department of Astronomy and Astrophysics, Chicago, IL USA}
\affiliation[c]{Fermi National Accelerator Laboratory, Theoretical Astrophysics Group, Batavia, IL USA}
\affiliation[d]{University of Chicago, Department of Physics, Chicago, IL USA}

\emailAdd{dhooper@fnal.gov}
\emailAdd{anireland@uchicago.edu}
\emailAdd{krnjaicg@fnal.gov}
\emailAdd{stebbins@fnal.gov}

\abstract{There is controversy surrounding the origin and evolution of our universe's largest supermassive black holes (SMBHs). In this study, we consider the possibility that some of these black holes formed from the direct collapse of primordial density perturbations. Since the mass of a primordial black hole is limited by the size of the cosmological horizon at the time of collapse, these SMBHs must form rather late, and are naively in conflict with constraints from CMB spectral distortions. These limits can be avoided, however, if the distribution of primordial curvature perturbations is highly non-Gaussian. After quantifying the departure from Gaussianity needed to evade these bounds, we explore a model of multi-field inflation --- a non-minimal, self-interacting curvaton model --- which has all the necessary ingredients to yield such dramatic non-Gaussianities. We leave the detailed model building and numerics to a future study, however, as our goal is to highlight the challenges associated with forming SMBHs from direct collapse and to identify features that a successful model would need to have. This study is particularly timely in light of recent observations of high-redshift massive galaxy candidates by the James Webb Space Telescope as well as evidence from the NANOGrav experiment for a stochastic gravitational wave background consistent with SMBH mergers.}

\begin{document}

\hfill{ \small FERMILAB-PUB-23-390-T}

\maketitle

\flushbottom

\section{Introduction}

\begin{figure}[t]
\centering
\includegraphics[width=0.77\textwidth]{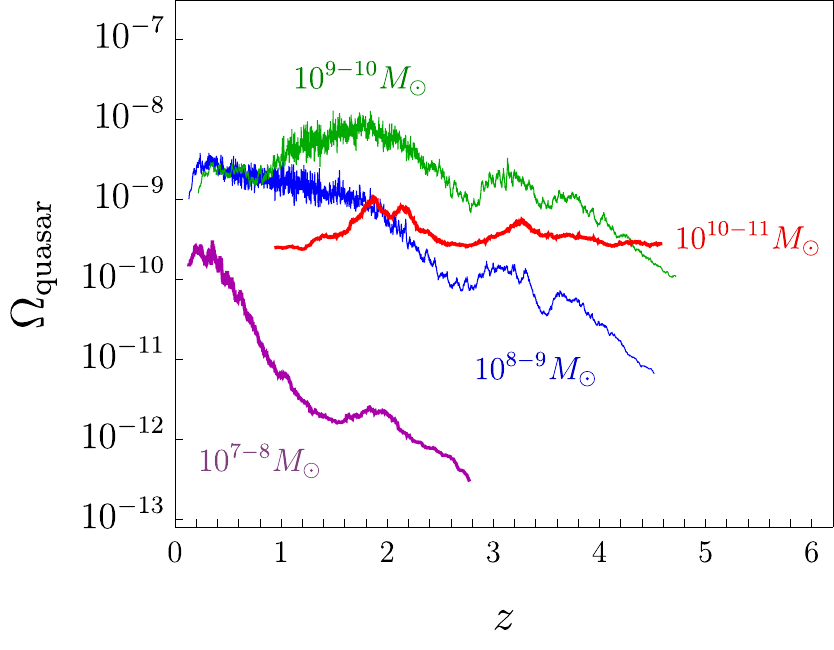}
\hspace{2mm}
\caption{Estimates of the quasar black hole comoving mass density in units of the $z=0$ critical density as a function of redshift, plotted for the four different mass classes indicated. Black hole masses and redshifts are taken from the DR7 SDSS quasar survey~\cite{Shen_2011}, which covers 20\% of the sky and the redshift range $0<z<5$. The results shown represent the moving average of 100 quasars sorted by redshift.}
\label{abundances}
\end{figure}

Supermassive black holes (SMBHs) are ubiquitous in our universe, being present in the centers of nearly all massive galaxies. Quasars,\footnote{Quasars are the most luminous active galactic nuclei (AGN).  In this paper, quasar refers to an AGN that is sufficiently luminous to appear in a quasar catalog such as SDSS DR7~\cite{Shen_2011}. While this definition is redshift-dependent, the most luminous quasars should be consistently present in the catalog up to the DR7 redshift limit of 5. In the text, the phrase ``quasar mass'' refers to the mass of the black hole that powers the quasar.} 
powered by black holes with with masses $M \sim 10^{8-10} \, M_{\odot}$, are also found in large numbers in the high-redshift universe. At present, over 170 quasars have been observed at $z>6$, with the most distant at $z=7.54$, and several hundred others at $z=5-6$~\cite{2018:Natur,2020MNRAS.494..789R,Willott:2009wv,Banados:2016mwo,2016ApJ...833..222J,DES:2017yqk,2019ApJ...883..183M,2018arXiv181011927Y,2019ApJ...884...30W}. Fig.~\ref{abundances} shows the quasar abundance as a function of redshift over several ranges of black hole mass.
While this magnitude limited quasar catalog is believed to be nearly complete out to $z \sim 5$, only a small fraction of SMBHs are, in fact, quasars.  A more complete census of SMBHs is possible in the local universe where one finds 
$\Omega_{\rm SMBH}(z = 0) \sim 10^{-6}$~\cite{Mutlu_Pakdil_2016}. This contrasts with the peak quasar mass density
of $\Omega_{\rm quasar}(z = 2)\sim10^{-8}$. While specific models for SMBH population evolution have been proposed~\cite{Shankar2012,Shen2020,Sicilia2022}, the limited available data leaves a great deal of uncertainty.

It is usually assumed that SMBHs grow over time from relatively low-mass seeds (possibly the remnants of Population III stars~\cite{Banik:2018}) through the process of accretion. The rate of mass accretion is Eddington limited to
\begin{equation}
	\dot{M}_{\rm Edd} \lesssim \frac{M_{\rm BH}}{\tau_S},
\end{equation}
where 
\begin{equation}
	\tau_\mathrm{S} = \frac{\varepsilon\,\sigma_\mathrm{T} }{4\pi\,G\,m_\mathrm{p}} \approx 45 \, \, {\rm Myr} \, \bigg(\frac{\varepsilon}{0.1}\bigg) \, 
\end{equation}
is the Salpeter time~\cite{1964ApJ...140..796S}, $\sigma_\mathrm{T} 
=8\pi\,\alpha^2/(3 m_{e}^2)$ is the Thomson cross section, $m_{p}$ is the proton mass, and $\varepsilon$ is the radiative efficiency. At the Eddington-limited rate, a $10^2\,M_{\odot}$ black hole seed would require $\sim0.8\,\mathrm{Gyr}$ to grow to $M\sim 10^{10}\,M_{\odot}$. Since $z \sim 6-7$ corresponds to only $\sim 0.7-0.9 \, {\rm Gyr}$ after the Big Bang, such a scenario would require these large high-redshift black holes to have grown at high accretion rates almost continuously throughout the first Gyr of our universe's history. Interestingly, there seems to be quite a significant population of quasars at $z\sim6$ to $7$ in this mass range~\cite{SDSS:2001emm,Willott:2009wv,Mortlock:2011va,2015Natur.518..512W}.

The continuous accretion required to explain this rapid growth contrasts with the intermittent accretion observed of SMBHs at lower redshifts. Further, from Fig.~\ref{abundances} we see that the most massive $M_{\rm BH} \gtrsim10^{10}\, M_{\odot}$ population has remained approximately constant since at least $z \sim 5$~\cite{Trakhtenbrot:2010hj}. This would require any growth in the number of SMBHs to be balanced by a decrease in the fraction of SMBHs that are actively quasars. From this perspective, it is surprising that so many highly-massive quasars have been observed at such high redshifts~\cite{Volonteri:2010wz,Haiman:2004ve,Shapiro:2004ud,Volonteri:2006ma,Tanaka:2008bv}. These observations prompt two intriguing questions:
\begin{enumerate}
	\item If these quasars grew from small black hole seeds, how did they come to be so massive on such a short timescale? 
	\item Why did their growth rate dramatically slow down during the subsequent $13 \, {\rm Gyr}$?
\end{enumerate}

Various solutions to the first question have been put forth, including an enhanced role of mergers as well as the possibility of super-Eddington accretion \cite{Volonteri:2021}. However frequent mergers would require more heavily clustered populations in the early universe, and further can have both positive and negative impacts on SMBH growth, since they can also kick SMBHs out of the material-rich centers of galaxies. A certain degree of super-Eddington growth is expected to occur in high-redshift galaxies containing large reservoirs of gas and efficient angular momentum transport due to turbulence. However as increased accretion produces powerful jets and outflows which drive material away, it is uncertain how long it can be sustained. Feedback effects from transient periods of super-Eddington growth are actually expected to impede SMBH growth within a few Myr \cite{Massonneau:2022}.

Regarding the second question, it has been suggested that the suppressed growth of the most massive black holes after the first Gyr could perhaps be attributed to galaxy-scale feedback. This, however, would require the $M$-$\sigma$ relation\footnote{The $M$-$\sigma$ relation is the observed correlation between the velocity dispersion of a stellar bulge $\sigma$ and the mass of the SMBH at its center.} to evolve with redshift and for the quasar luminosity function to steepen at the highest values~\cite{Netzer:2002vw,Natarajan:2008ks}. Alternatively, it has been proposed that there might be a maximum mass that black holes can reach through accretion, resulting from the fragmentation of the accretion disks that could have otherwise facilitated rapid black hole growth~\cite{2016ApJ...828..110I}. Despite these suggestions, there remain many open questions concerning the origin and evolution of our universe's most massive black holes.

In this study, we take these questions to motivate another possibility: that our universe's most massive black holes did not acquire most of their mass through accretion, but are instead predominantly primordial in origin\footnote{Primordial black holes (PBHs) have previously been considered as SMBH seeds \cite{Kawasaki:2012}. Unlike population III stars, which do not form until $z\sim 30-20$, PBH seeds form much earlier, and so the timing problem is allayed. Refs.~\cite{Bernal:2017,Smith:2017} have argued that establishing a population of PBHs with mass $10^2-10^6 M_\odot$ by $z \sim 20$ would be sufficient to seed even the most massive SMBHs. Here we consider the more exotic possibility of SMBHs from direct collapse, with little growth through accretion required.}. Unlike the small primordial black holes (PBH) typically considered as dark matter candidates, these massive objects would have formed at ``late'' times, as governed by the size of the cosmological horizon. During the radiation dominated era, the horizon contains the following amount of energy: 
\begin{equation}\label{horizonmass}
	M_{H} = \frac{4\pi}{3}  \rho R_H^3 \approx 3 \times 10^9 \, M_{\odot} \, \bigg(\frac{10 \, {\rm keV}}{T}\bigg)^2 \bigg(\frac{3.36}{g_{\star}(T)}\bigg)^{1/2} \,,
\end{equation}
where $R_H = H^{-1}$ is the size of the horizon, $\rho = \pi^2 g_\star(T) T^4/30$ is the radiation density, and $g_\star$ is the number of relativistic species at temperature $T$. When a sufficiently large density fluctuation collapses to form a PBH,\footnote{There exist other PBH formation mechanisms beyond the collapse of overdensities seeded by inflation. For example, PBH could form during first order cosmological phase transitions \cite{Liu:2021} or due to the collapse of supercritical vacuum bubbles nucleated during inflation \cite{Garriga:2015}. The prospect of primordial SMBHs from these mechanisms is discussed in Ref.~\cite{Davoudiasl:2021} and Ref.~\cite{Huang:2023}, respectively.} the mass of the resulting black hole is typically an order one fraction of the horizon mass, $M_{\rm BH} \simeq \gamma \, M_{H}$, where $\gamma \sim 0.2$ quantifies the efficiency of collapse~\cite{Carr:1975qj}. From Eq.~(\ref{horizonmass}), we conclude that the PBHs in the mass range of interest here form after the start of Big Bang nucleosynthesis ($T\sim {\rm MeV}$ corresponds to $M_H \sim 10^5 \, M_{\odot}$) but well before the onset of matter domination or recombination ($T\sim {\rm eV}$ corresponds to $M_H\sim 3 \times 10^{17} \, M_{\odot}$). 

Many inflationary models enable PBHs to form efficiently~\cite{Leach:2001}, including those in which the inflaton 
undergoes a period of ultra-slow-roll~\cite{Bellido:2017,Germani:2017,Ballesteros:2017,Hertzberg:2017}, 
or whose potential features localized bumps, dips, or steps on small scales~\cite{Mishra:2019,Inomata:2021,Cai:2022}. More generically, a local enhancement of the power spectrum $\mathcal{P}_\zeta$ requires a deviation from slow-roll evolution (see for example, Refs.~\cite{Pattison:2017,Biagetti:2018,Ezquiaga:2019,Pattison:2021,Cai:2021,Garcia-Bellido:1996,Clesse:2015,Brown:2017,Palma:2020,Fumagalli:2020,Braglia:2020,Pi:2021,Geller:2022,Ozsoy:2023,Cai:2023}). Regardless of the mechanism, PBH formation requires a significant enhancement in the amplitude of the primordial power spectrum. Assuming Gaussian statistics, this amplitude must be $\mathcal{P}_\zeta(k_{\rm BH}) \sim 10^{-2}$, seven orders of magnitude larger than observed on large scales, $\mathcal{P}_\zeta(k_{\rm CMB}) \simeq 2.1\times 10^{-9}$~\cite{Planck:2018}. Such an amplification inevitably leads to  large CMB spectral distortions. In particular, for scales $k_{\rm BH} \lsim 10^5 \, {\rm Mpc}^{-1}$, corresponding to black holes in the mass range $M_{\rm BH} \gtrsim 10^3 M_\odot$, the predicted spectral distortions are in strong conflict with COBE/FIRAS measurements~\cite{Mather:1993,Fixsen:1996}. 

In principle, PBHs can form from smaller peaks in the power spectrum if the tail of the $\zeta$ distribution is sufficiently non-Gaussian. Of course, observations on large scales and measurements of the non-linearity parameters $f_{\rm NL}$ and $g_{\rm NL}$ seem to indicate that $\zeta$ is very nearly Gaussian on CMB scales~\cite{Planck:2018IX,Planck:2018X}. However the scales on which PBHs are formed can be disconnected from CMB scales, and so these bounds need not apply.


In this paper, we  quantify  the degree of non-Gaussianity that would be required to viably produce primordial SMBHs and present a model of inflation that contains the necessary features. We begin in Sec.~\ref{PBHformation} by reviewing the calculation of the PBH abundance in the Press-Schechter formalism. In Sec.~\ref{spectraldistortions}, we examine how measurements of spectral distortions constrain the primordial power spectrum, and demonstrate that the appreciable formation of SMBHs from Gaussian density fluctuations is excluded on the basis of these constraints. Sec.~\ref{departures} considers departures from Gaussianity, and quantifies the heaviness of the distribution needed to evade these bounds.  In Sec.~\ref{standardcurvaton}, we review the calculation of the curvature perturbation and its statistics in the standard curvaton scenario, which has been shown to be capable of producing appreciable non-Gaussianity. This model, however, cannot viably produce a significant abundance of primordial SMBHs, so in Sec.~\ref{nonstandard} we introduce self-interactions to augment the non-Gaussianity and speculate on the maximum PBH abundance in this model. We conclude in Sec.~\ref{conclusions} with a discussion of our results and some comments on directions for future investigations.

\section{Primordial Black Holes from Gaussian Perturbations}\label{PBHformation}

The initial abundance of PBHs is quantified in terms of their mass fraction at the time of formation:
\begin{equation}
	\beta \equiv \frac{\rho_{\rm BH}}{\rho_{\rm tot}} \,.
\end{equation}
For PBHs that formed during radiation domination, this quantity is related to the fractional energy density in black holes today, $\Omega_{\rm BH} \simeq \beta \, (1+z_f) \,\Omega_{R}$, where $z_f$ is the redshift at the time of black hole formation. Using Eq.~(\ref{horizonmass}) and entropy conservation, this can be written more conveniently in terms of the initial PBH mass as:
\begin{equation}\label{OmegaBH}
    \Omega_{\rm BH} \simeq  10^{-6} \, \left(\frac{\beta}{10^{-10}}\right) \bigg(\frac{\gamma}{0.2}\bigg)^{1/2} \left(\frac{10^8 \, M_{\odot}}{M_{\rm BH}}\right)^{1/2} \left( \frac{g_{\star,s}(T_f)}{3.91} \right)^{1/3} \left( \frac{3.36}{g_{\star}(T_f)} \right)^{1/4},
\end{equation}
where $T_f$ is the temperature at the time of black hole formation.

In the standard\footnote{One can also calculate the black hole abundance using peak theory~\cite{Bardeen:1985,Green:2004,Young:2020,Yoo:2020}. Unlike Press-Schechter, where the overdensity must simply exceed the threshold, peak theory further demands that it be a local maximum. This formalism has been demonstrated to be more appropriate when perturbations exist on multiple scales. As we consider a sharply peaked spectrum, the simpler Press-Schechter prescription suffices for our purposes. A comprehensive comparison can be found in \cite{Gow:2020}, which demonstrated that differences are minimal so long as the window function is implemented properly.} treatment based on the Press-Schechter formalism~\cite{Press:1973}, $\beta$ is computed by integrating the probability distribution $P_\delta$ for the coarse-grained density contrast $\delta = \delta \rho/\bar{\rho}$ over all values greater than the critical threshold for collapse, $\delta_c$:
\begin{equation}\label{betaeq}
    \beta \simeq 2 \int_{\delta_c}^\infty d\delta \, P_\delta[\delta] \,.
\end{equation}
The factor of $2$ is customarily introduced\footnote{It is unclear whether this factor should still be included when considering asymmetric probability distribution functions, as in the case of non-Gaussianities. We retain it nevertheless since this is an ultimately inconsequential $\mathcal{O}(1)$ effect.} to compensate for the undercounting that otherwise arises~\cite{Press:1973}. This prescription has the benefit of having an intuitive interpretation --- we are summing over the fraction of regions with a sufficient overdensity to collapse to form a PBH --- and a simple implementation. One issue concerns the uncertainty\footnote{It has been demonstrated that a more precise quantity which does away with this uncertainty is the compaction function $\mathcal{C}$, defined as twice the local mass excess over the comoving areal radius~\cite{Shibata:1999}. The value of the compaction at its peak is interpreted as the threshold. See also Ref.~\cite{Escriva:2019}.} of the collapse threshold, which should in principle depend on the overdensity profile\footnote{Since the overdensity profile is modified in the presence of non-Gaussianities, these can also change the threshold for collapse~\cite{Atal:2019a,Atal:2019b,Yoo:2019,Kehagias:2019}. Given the uncertainties involved in these calculations, we take a conservative position and determine $\delta_c$ using Eq.~(\ref{threshold}).}. Assuming a spherical profile, we expect a perturbation to be able to collapse during radiation domination if the size of the overdensity at maximum expansion exceeds the Jeans length. This led to Carr's original estimate of $\delta_c = c_s^2 = 1/3$ \cite{Carr:1975qj}, where $c_s$ is the sound speed of density perturbations. A more careful treatment that is applicable for an arbitrary equation of state\footnote{It was argued in \cite{Carr:2019} that the softening of the equation of state at electron-positron annihilation should lead to an enhancement in the formation of PBHs with $M_{\rm BH} \sim 10^6 M_\odot$. However \cite{Musco:2023} pointed out that due to the onset of neutrino free-streaming, overdensities are reduced at this time, leading to an overall suppression of PBH formation.}, $w$, finds~\cite{Harada:2013}:
\begin{equation}\label{threshold}
    \delta_c = \frac{3(1+w)}{5+3w} \sin^2 \left( \frac{\pi \sqrt{w}}{1+3w} \right) \,,
\end{equation}
which has been shown to faithfully replicate the results of numerical simulations and yields $\delta_c = 0.414$ during radiation domination. We adopt this value throughout our analysis. 

While the coarse-grained density contrast is the proper object to consider when computing the probability of PBH formation, it is often convenient to work directly with the curvature perturbation $\zeta$, since its statistics are more easily computed from underlying inflationary models. When large perturbations exist only on one scale, as is the case for the sharply peaked power spectra we consider, only a minimal amount of error is incurred by making this approximation. On superhorizon scales, $\zeta$ is related to the density contrast field by~\cite{Ferrante:2022}
\begin{equation}\label{densitycontrast}
    \delta = - \frac{2(1+w)}{5+3w} \left( \frac{1}{aH} \right)^2 e^{-2 \zeta} \left( \nabla^2 \zeta + \frac{1}{2} (\partial_i \zeta)^2 \right) \,.
\end{equation}
Working to linear\footnote{A certain degree of non-Gaussianity inevitably arises in the density contrast field due to this nonlinear relation. This implies that even if the statistics of $\zeta$ were perfectly Gaussian, those of $\delta$ would not be. See, for example, Ref.~\cite{Ferrante:2022} for discussion.} order in radiation domination, this simplifies to the Fourier space relation:
\begin{equation}\label{deltaR}
    \delta_k \simeq \frac{4}{9} \left( \frac{k}{aH} \right)^2 \zeta_k \,,
\end{equation}
which implies that their power spectra, defined for arbitrary Fourier variable $f_k$ via the 2-point function as $\expval{f_k f_{k'}} = \frac{2\pi^2}{k^3} \mathcal{P}_f(k) \delta^{(3)}(\vec{k}+ \vec{k}')$, are related as
\begin{equation}
    \mathcal{P}_\delta(k) \simeq \frac{16}{81} \left( \frac{k}{aH} \right)^4 \mathcal{P}_\zeta(k) \,.
\end{equation}
Note that at the time of horizon crossing $k \simeq aH$, when a perturbation can collapse to form a black hole,\footnote{There is additional uncertainty in the calculation of PBH abundance coming from non-linear effects around the time of horizon crossing. See Ref.~\cite{DeLuca:2023}.} the density contrast and curvature perturbation are linearly related, $\delta \simeq \frac{4}{9} \zeta$. We can then assume that peaks in $\delta$ also correspond to peaks in $\zeta$, and work directly with the curvature perturbation. In terms of $\zeta$, the initial abundance in the Press-Schechter formalism reads:
\begin{equation}\label{betaestimate}
    \beta \simeq 2 \int_{\zeta_c}^\infty d\zeta \, P_\zeta[\zeta] \,,
\end{equation}
where the collapse threshold $\zeta_c \simeq \frac{9}{4} \delta_c$ follows from $\sigma_\delta^2 \simeq \frac{16}{81} \sigma_\zeta^2$, since $\mathcal{P} \sim \sigma^2$ for a sharply peaked spectrum. More precisely, the variance of $\zeta$ smoothed on the scale $R \simeq (aH)^{-1} \simeq k^{-1}$ can be computed from the power spectrum as \cite{Young:2014}
\begin{equation}\label{variance}
    \sigma_\zeta^2(R) \equiv \expval{\zeta}^2_R = \int_0^\infty \frac{dk}{k} \tilde{W}^2(k,R) \mathcal{P}_\zeta(k) \,,
\end{equation}
where $\tilde{W}(k,R)$ is the Fourier transform of the (real space) window function used to coarse-grain $\delta$. It is unclear what functional form for the window function most accurately reproduces the actual relation between the PBH abundance and power spectrum, but popular choices in the literature include the (volume normalized) Gaussian, as well as real and $k$-space top hats. For simplicity, we use the former, $\tilde{W}(k,R) = \exp \left( - k^2 R^2/2 \right)$, but see \cite{Ando:2018,Young:2019} for a discussion of the resultant uncertainties.

Consider the case of a Gaussian distribution $P_\zeta = P_\zeta^{G}$, where
\begin{equation}\label{pG}
    P_\zeta^{G} = \frac{1}{\sqrt{2\pi} \sigma_\zeta} e^{- \zeta^2/2 \sigma_\zeta^2} \,.
\end{equation}
In this case the mass fraction at formation evaluates to
\begin{equation}
    \beta = \text{erfc} \left( \frac{\zeta_c}{\sqrt{2} \sigma_\zeta} \right) \simeq \sqrt{\frac{2}{\pi}} \frac{\sigma_\zeta}{\zeta_c} \exp \left(- \frac{\zeta_c^2}{2 \sigma_\zeta^2} \right) \,,
\end{equation}
where $\text{erfc}$ is the complimentary error function and the second approximation holds for $\zeta_c \gg \sigma_\zeta$, which is generically true for all cases of physical interest. Since the threshold is fixed, the initial abundance is determined solely by the variance of the power spectrum. In order to have $\beta = 10^{-20}$, we see we need $\sigma_\zeta^2 \simeq 0.01$, corresponding to a peak in the power spectrum of $\mathcal{P}_\zeta \sim 0.01$, which is 7 orders of magnitude greater than the value measured on CMB scales. Since the amplification of $\mathcal{P}_\zeta$ needed for PBH formation depends on $\beta$ only logarithmically, this degree of enhancement is a generic requirement for any non-vanishing initial abundance.


\section{Spectral Distortions}\label{spectraldistortions}

The small scales relevant for primordial SMBH formation are well below the angular resolution probed by current CMB measurements. Nevertheless, inhomogeneities on these scales will generate isotropic deviations from the usual blackbody spectrum~\cite{Sunyaev:1970,Zeldovich:1969,Illarionov:1975,Hu:1992,Chluba:2012}. These deviations are known as spectral distortions and in this context
it is useful to distinguish between three characteristic redshift intervals: 
\begin{itemize}
\item {{\bf Thermalization era (}$\bm{ z > 2 \times 10^6})$}: At high redshifts, Compton scattering $\gamma e \to \gamma e$, double Compton scattering $\gamma e \to \gamma \gamma e$, and Bremsstrahlung $e p \to ep \gamma$ maintain a blackbody spectrum for the photons, and spectral 
distortions are exponentially suppressed.

\item  $\bm{ \mu}$-{\bf era}~$(\bm{2 \times 10^5 < z < 2 \times 10^6})$:
During this era, photon number changing processes, double Compton scattering  and Bremsstrahlung,
become ineffective at maintaining a blackbody spectrum. Compton scattering, however, continues to redistribute photon energies to maintain a Bose-Einstein distribution, parameterized by both a temperature, $T$, and a chemical potential, $\mu$.  A $\mu$-distortion refers to a Bose-Einstein distribution with $\mu\ne0$.

\item { $\bm y$-{\bf  era (}$\bm{ z < 2 \times 10^5})$}:
Compton scattering becomes ineffective at redistributing photon energies during this era, so there are no processes to maintain a Bose-Einstein distribution. Spectral distortions that are generated at these redshifts are characterized by a departure from an equilibrium distribution, and often yield so-called $y$-distortions.\footnote{Technically, the division between the two types of spectral distortions is not entirely unambiguous, and inhomogeneities dissipating around $z \sim 5 \times 10^4$ can give rise to distortions of an intermediate type, whose shape is not simply the sum of $\mu$- and $y$-type distortions~\cite{Khatri:2012}.} 

\end{itemize}

A spectrum with a (positive) $y$-distortion can be expressed as an average of blackbodies with slightly different temperatures~\cite{Chluba2004,Stebbins:2007,Chluba:2011,Pitrou:2014}. An average of blackbodies with a mean temperature $\bar{T}$ and variance $\bar{T}^2\,\Delta$ will (for $\Delta\ll1$) be a $y$-distorted blackbody characterized by the temperature $T=\bar{T}\,(1+\Delta^2)$ and the parameter $y=\Delta^2/2$. $y$-type distortions can be generated through a variety of mechanisms, including the Compton scattering of CMB photons with a population of electrons with a different temperature. In this paper, we will be interested in $y$-distortions that are generated through photon diffusion. Limits on such spectral distortions allow us to constrain the amplitude of inhomogeneities on very small scales~\cite{Carr1993}. 


$\mu$- and $y$-type spectral distortions are traditionally quantified in terms of the parameters $\mu$ and $y$, which are related to the fractional increase in energy per photon (relative to a blackbody spectrum with the same number density of photons)~\cite{Sunyaev:1970,Chluba:2012}:
\begin{equation}
    \mu \simeq 1.4 \, \frac{\Delta \rho_\gamma}{\rho_\gamma} \qquad\textrm{and}\qquad
    y \simeq 0.25 \, \frac{\Delta \rho_\gamma}{\rho_\gamma}\ .
\end{equation}
By introducing $k$-space window functions accounting for the effects of thermalization and dissipation, $\mu$ and $y$ can be approximately calculated
from the spectrum of density perturbations, $P_\zeta$~\cite{Chluba:2011,Chluba:2012}:
\begin{align}
   & \mu \simeq 2.2 \int_{k_{\rm min}}^\infty \frac{dk}{k} \mathcal{P}_\zeta(k) \left[ \exp \left( - \frac{k}{5400\, \text{Mpc}^{-1}} \right) - \exp \left( - \left[ \frac{k}{31.6 \, \text{Mpc}^{-1}}\right]^2 \right) \right] \,,\\[10pt] 
    & y \simeq 0.4 \int_{k_{\rm min}}^\infty \frac{dk}{k} \mathcal{P}_\zeta(k) \exp \left( - \left[ \frac{k}{31.6 \, \text{Mpc}^{-1}}\right]^2 \right) \,,
\end{align}
where $k_{\rm min} = 1\, \text{Mpc}^{-1}$. 

The strongest existing constraints on spectral distortions come from the COBE/FIRAS instrument, which restricts $|\mu| \lesssim 9.0 \times 10^{-5}$ and $|y| \lesssim 1.5 \times 10^{-5}$ at the 95\% C.L.~\cite{Mather:1993,Fixsen:1996}. The standard cosmological model predicts spectral distortions of $\mu\sim2\times10^{-8}$ and $y\sim10^{-6}$~\cite{Chluba:2016}, consistent with current limits. The models of interest in this study have enhanced $P_\zeta$, and thus enhanced $\mu$ and $y$, which can be constrained by CMB measurements.

For concreteness, consider the case of a power spectrum that is sharply peaked at a single scale, $k_{\rm BH}$:
\begin{equation}\label{deltaPR}
    \mathcal{P}_\zeta(k) = \sigma_\zeta^2 \, k \, \delta(k - k_{\rm BH}).
\end{equation}
Using the delta-function to perform the integral, $\mu$ and $y$ become functions of $k_{\rm BH}$ and $\sigma_\zeta^2$ alone. From the horizon crossing condition $k = a H$, and fact that $H = 1.66 \sqrt{g_{\star} (T)} T^2/M_{\rm Pl}$ during radiation domination, we can relate the wavenumber to the temperature,
\begin{equation}
    k_{\rm BH} = 92 \, {\rm Mpc}^{-1} \, \left(\frac{T}{10 \, {\rm keV}}\right) \, \left(\frac{g_{\star}(T)}{3.36}\right)^{1/2} \, \left(\frac{3.91}{g_{\star, S}(T)}\right)^{1/3} \,,
\end{equation}
which can then be related to the horizon mass using Eq.~(\ref{horizonmass}) to estimate the mass of the resulting black hole:
\begin{equation}\label{MPBH}
    M_{\rm BH} \simeq 5 \times 10^{8} \, M_{\odot} \, \left( \frac{92 \, \text{Mpc}^{-1}}{k_{\rm BH}} \right)^2 \left(\frac{\gamma}{0.2}\right) \left(\frac{g_{\star}(T)}{3.36}\right)^{1/2} \left(\frac{3.91}{g_{\star, S}(T)}\right)^{2/3} \,.
\end{equation}
For the sharply peaked spectrum of Eq.~(\ref{deltaPR}), adopting $\gamma=0.2$, and for black holes in the mass range of interest, the $\mu$- and $y$-parameters can be written as
\begin{align}\label{SDestimates}
    & \mu \simeq 2.2 \sigma_\zeta^2 \, \Bigg\{ \exp \left[ - \left( \frac{1.5 \times 10^5 \, M_\odot}{M_{\rm BH}} \right)^{1/2}  \right] 
    - \exp \left[ - \left( \frac{4.5 \times 10^9 \, M_\odot}{M_{\rm BH}} \right) \right] \Bigg\} \,, \\[6pt]
    & y \simeq 0.4 \sigma_\zeta^2 \, \exp \left[ - \left( \frac{4.5 \times 10^9 \, M_\odot}{M_{\rm BH}} \right) \right] \,.
    \label{SDestimates2}
\end{align}
Note that these results should also hold, for example, in the case of a log-normal spectrum of sufficiently narrow width.

\begin{figure}[t]
\centering
\includegraphics[width=0.86\textwidth]{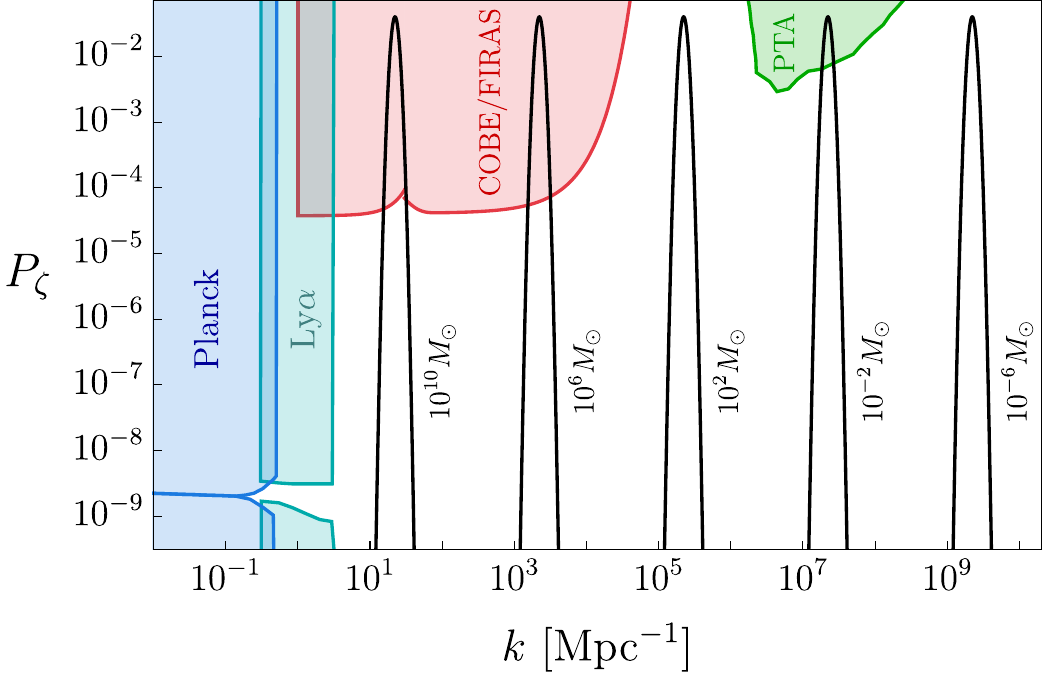}
\hspace{2mm}
\caption{Constraints on the primordial power spectrum $\mathcal{P}_\zeta$ \cite{Green:2020} coming from CMB temperature anisotropies (dark blue) \cite{Planck:2018X}, Lyman-$\alpha$ forest (light blue) \cite{Bird:2010}, CMB spectral distortions (red) \cite{Mather:1993,Fixsen:1996}, and pulsar timing arrays (green) \cite{Byrnes:2018}. The cusp in the COBE/FIRAS excluded region signifies the wavenumber where constraints from $\mu$- and $y$-type distortions are equally restrictive. Overlaid are illustrative sharply peaked log-normal power spectra resulting in the formation of PBHs with $M_{\rm BH} = 10^{10}, 10^6, 10^2, 10^{-2},\, \text{and} \,10^{-6} M_\odot$ and an initial abundance of $\beta = 10^{-20}$, assuming Gaussian statistics for $\zeta$.
}
\label{SDconstraints}
\end{figure}

In Eqs.~(\ref{SDestimates}) and~(\ref{SDestimates2}), we can identify the impact of the various eras described earlier in this section. In particular, for $M_{\rm BH} \ll 1.5 \times 10^5 \, M_{\odot}$ (corresponding to $k_{\rm BH} \gg5400/\textrm{Mpc}$), the black holes are formed during the thermalization era, and both $\mu$- and $y$-type spectral distortions are suppressed. For $1.5 \times 10^5 \, M_{\odot} \ll M_{\rm BH} \ll   4.5 \times 10^9 \, M_{\odot}$ ($31.6/\textrm{Mpc}\lesssim k_{\rm BH }\lesssim5400/\textrm{Mpc}$), the black holes are forming during the $\mu$-era, leading primarily to $\mu$-type spectral distortions. Larger black holes form later, yielding primarily $y$-type spectral distortions.

We can use Eqs.~(\ref{SDestimates}) and~(\ref{SDestimates2}) to quickly estimate whether a scenario featuring primordial SMBHs is consistent with spectral distortion constraints. Recall that, for the case of Gaussian statistics, a value of $\sigma_\zeta^2 \gsim 10^{-2}$ is required in order to obtain a non-negligible abundance of PBHs. For such large values of $\sigma_\zeta^2$, spectral distortions exclude all black holes masses $M_{\rm BH} \gtrsim \text{few} \times 10^3 M_\odot$. This conclusion is consistent with Fig.~\ref{SDconstraints}, where we compare the bounds from COBE/FIRAS with the power spectra predicted for several values of $M_{\rm BH}$. Therefore, the existence of a non-negligible abundance of primordial SMBHs requires the presence of significant non-Gaussianities in the distribution of the primordial curvature perturbations.


\section{Departures from Gaussianity}\label{departures}

For Gaussian density perturbations, constraints from spectral distortions severely limit the abundance of primordial SMBHs that could have formed in the early universe. In this case, the variance must be small in order to limit spectral distortions, while a large variance is required to generate a non-negligible abundance of PBHs. This tension could be resolved, however, if the distribution of curvature perturbations features a heavier tail than that of a Gaussian. Such non-Gaussianities thus potentially allow primordial SMBHs to form\footnote{As noted in \cite{Gow:2023}, large curvature perturbations in models with heavy-tailed distributions are often type~II. Such perturbations have been demonstrated to form PBHs \cite{Kopp:2010}, however collapse in the type II case --- in particular the mass of the resultant PBH --- is not completely understood.} without necessarily violating spectral distortion constraints.\footnote{Ref.~\cite{Nakama:2016} proposes an alternative way to obtain the extreme statistics needed to form primordial SMBHs compatible with spectral distortion bounds. Their multi-field model results in effectively two different inflationary histories for the casually disconnected Hubble patches, with a subdominant fraction experiencing more expansion. By the $\delta N$ formalism this equates to large curvature perturbations, such that these patches collapse to form PBH. Since this occurs only in a tiny fraction of patches, there is no observable generation of spectral distortions.}

Fortunately, the distribution of primordial density perturbations is generically predicted to be non-Gaussian. Firstly, there is intrinsic non-Gaussianity that arises from the non-linear mapping between the curvature perturbation $\zeta$ and the density contrast $\delta$, as can be seen in Eq.~(\ref{densitycontrast}). Thus, even if the probability distribution function for $\zeta$ were exactly Gaussian, the distribution in $\delta$ would not be. Secondly, and more crucially, large departures from Gaussianity are generically found in models which produce a local enhancement in the primordial power spectrum~\cite{Ferrante:2022}. 

To quantify the degree of non-Gaussianity that would be required to generate primordial SMBHs without violating spectral distortion constraints, it is instructive to consider a class of probability distribution functions of the form~\cite{Nakama:2016}
\begin{equation}\label{pn}
    P^{(n)}_\delta = \frac{1}{2 \sqrt{2} \, \sigma_0 \, \Gamma\left(1 + \frac{1}{n} \right) } \exp \left[ - \left( \frac{|\delta|}{\sqrt{2} \sigma_0} \right)^{\! n} \,\right] \,,
\end{equation}
where $n$ parameterizes the heaviness of the distribution's tail. The variance of the density contrast is set by the second moment of the distribution:
\begin{equation}
    \sigma_\delta^2(\sigma_0) \equiv \int_{-\infty}^{\infty} d \delta \, \delta^2 \, P^{(n)}_\delta = \frac{ \,\,
\sigma_0^2 \, \Gamma \left(1 + \frac{3}{n} \right)
    }{3 \Gamma\left(1 + \frac{1}{n}  \right)}~.
\end{equation}
Note that for $n=2$, this reduces to the Gaussian form of Eq.~(\ref{pG}) with $\sigma_0^2 = \sigma_\delta^2$.
For $n=1$ the tail falls off exponentially, while for $0 < n < 1$ it falls off even more slowly; we will refer to this class of distributions with $n<1$ as ``heavy-tailed.''\footnote{Formally, the probability distribution $P_x$ of a random variable $x$ is said to be ``heavy'' if its tail is not exponentially bounded, $\lim_{x \rightarrow \infty} e^{\lambda x} \bar{F}(x) = \infty \, \forall \, \lambda>0$, where $\bar{F}(x) = \int_x^\infty dx' \, P_x(x')$.}

\begin{figure}[t!]
\centering
\hspace{-1cm}
\includegraphics[width=0.49\textwidth]{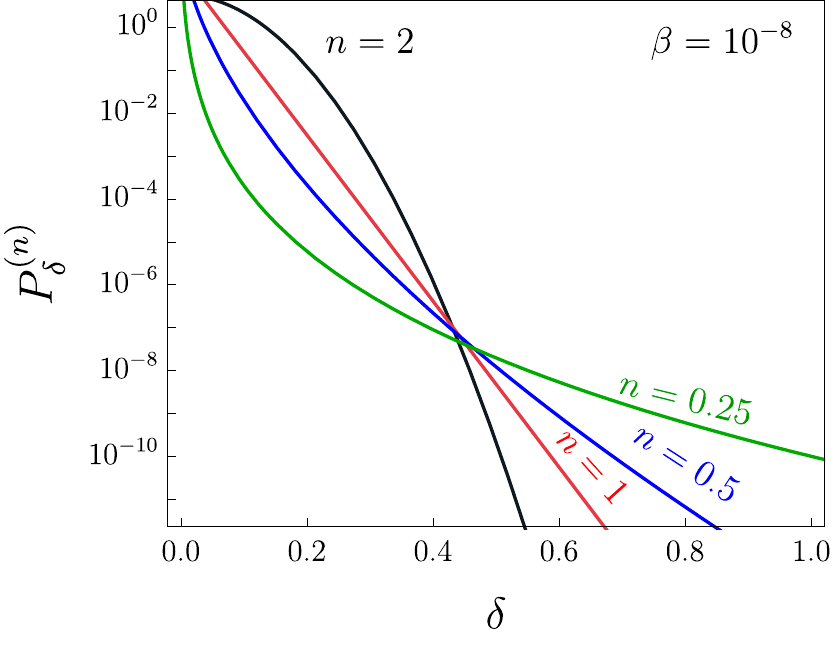}~~
\includegraphics[width=0.49\textwidth]{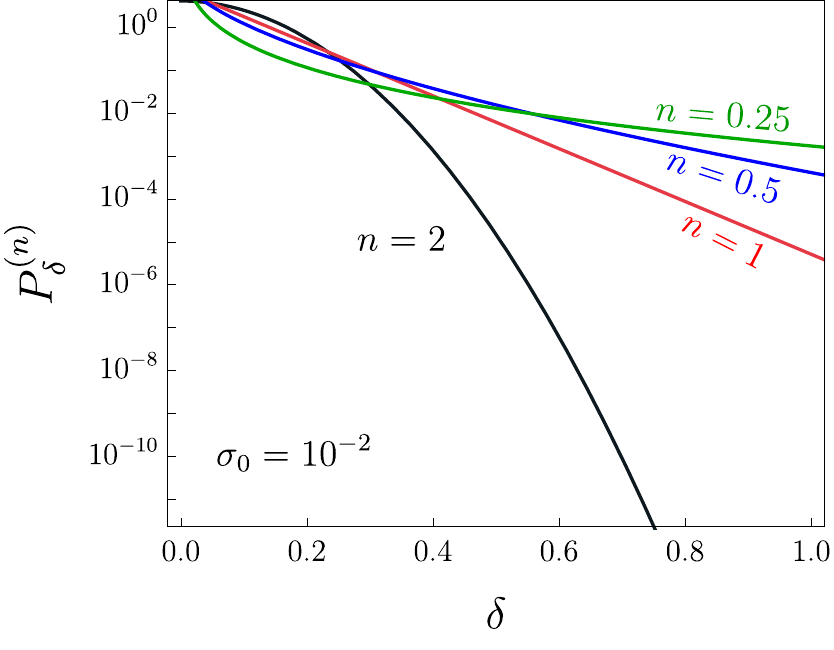}
\caption{Probability distribution functions Eq.~\ref{pn} for various choices of $n$ at fixed $\beta$ (left) and fixed variance $\sigma^2$ (right). We see that smaller $n$ corresponds to heavier-tailed distributions.
}
\label{pdf}
\end{figure}

\begin{figure}[t]
\centering
\includegraphics[width=0.77\textwidth]{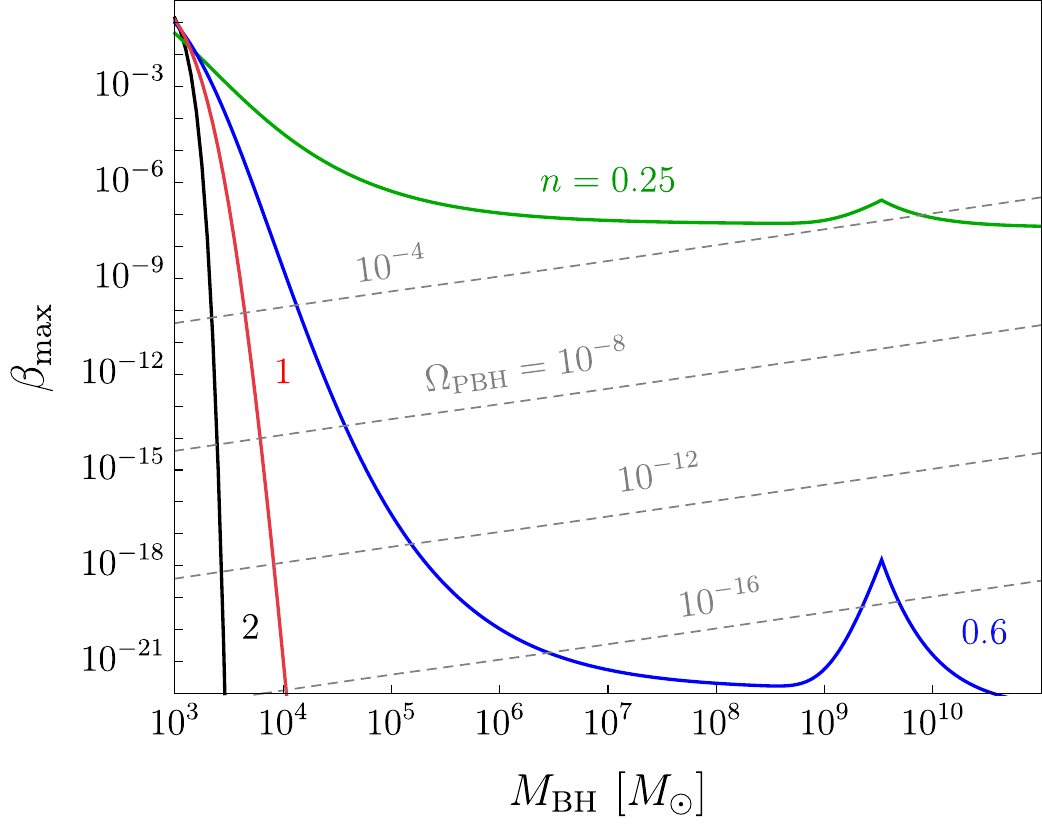}
\hspace{2mm}
\caption{The maximum primordial black hole mass fraction at formation $\beta_{\rm max}$ as a function of mass $M_{\rm BH}$ for a value of the variance $\sigma_\zeta^2$ that saturates the spectral distortion constraints from COBE/FIRAS, as estimated according to Eq.~(\ref{SDestimates}). We assume the distribution function given in Eq.~(\ref{pn}) and consider Gaussian ($n=2$, black), exponential ($n=1$, red), and power law ($n=0.6$ blue, $n = 0.25$ green) behavior in the tail. Note that the cusps which appear near $M_{\rm BH} \sim 3 \times 10^{9}\, M_\odot$ correspond to the value of $M_{\rm BH}$ at which $\mu$- and $y$-type spectral distortions are equally restrictive. Contours of constant $\Omega_{\rm BH}$, as computed using Eq.~(\ref{OmegaBH}), are shown in dashed gray.}
\label{betamax}
\end{figure}

In Fig.~\ref{pdf}, we show the shape of this class of probability distributions for various choices of $n$.
As expected, we see that smaller $n$ gives rise to heavier tails. This raises the question of how small $n$ must be in order to efficiently form primordial SMBHs while 
keeping the peak of the power spectrum within bounds of spectral distortions, $\mathcal{P}_\zeta \lesssim 10^{-4}$. In Fig.~\ref{betamax}, we plot the maximum PBH mass fraction at formation $\beta_{\rm max}$, for a variance that saturates the spectral distortion constraints from COBE/FIRAS. Note that to generate a present day abundance of $\Omega_{\rm BH} \gtrsim 10^{-20}$
with $M_{\rm BH} \lsim 10^{11} M_\odot$, 
we need a heavy tail with $n \lesssim 0.6$.

We now turn to the question of what inflationary models could yield such dramatic departures from Gaussianity. In the context of single-field inflation, Refs.~\cite{Namjoo:2012,Chen:2013a,Chen:2013b,Cai:2018} perturbatively studied the local non-Gaussianity that arises in models which deviate from the slow-roll attractor, as in ultra-slow-roll inflation. Going beyond perturbation theory, Refs.~\cite{Cai:2021,Cai:2022} used the $\delta N$ formalism \cite{Salopek:1990,Starobinsky:1982,Sasaki:1995,Sasaki:1998,Lyth:2004,Lyth:2005} to compute the non-perturbative distribution of curvature perturbations.\footnote{Ref.~\cite{Gow:2022} presents another non-perturbative method for calculating the PBH abundance in models where $\zeta$ is related to a Gaussian reference variable $\zeta_G$ by a ``generalized local transformation''.} For inflaton potentials with a small step or bump-like feature that induces a period of off-attractor behavior, these studies found that the tail of the distribution can become exponential, without inducing any significant non-Gaussianities in the perturbative regime~\cite{Cai:2022}. This result highlights the fact that perturbative measures of non-Gaussianity are not generally adequate to describe the large, rare fluctuations that lead to PBH formation. Finally, many of the mechanisms for enhancing local curvature perturbations rely on a temporary reduction in the inflaton's velocity. When the slow-roll classical drift vanishes, the field dynamics can receive large corrections from quantum diffusion, and the stochastic inflationary formalism~\cite{Starobinsky:1982,Starobinsky:1986} may be necessary for a proper description of the dynamics. Combining this with the $\delta N$ formalism~\cite{Vennin:2015}, a number of studies~\cite{Pattison:2017,Ezquiaga:2019,Pattison:2021,Tada:2021,Firouzjahi:2018,Achucarro:2021} have found that prominent exponential tails arise generically from quantum diffusion.  

While many single-field models have been found to yield exponential tails \cite{Pi:2023}, there is currently no known model which generates a heavier-tailed distribution.\footnote{Ref.~\cite{Atal:2020yic} interprets the NANOGrav signal as evidence of PBH mergers with $M_{\rm BH} \sim 10^{11}-10^{12} M_\odot$, and claims that $\mu$-distortion constraints can be overcome for sufficiently non-Gaussian single-field models. However they make no reference to $y$-type distortions, which are more constraining for this mass range, and which we have verified rule out this single-field scenario for any non-negligible abundance.}
%
However, as shown in Fig.~\ref{betamax}, a heavy-tailed $P_\delta \sim \exp\left( -|\delta|^n \right)$ with $n \lesssim 0.6$ is needed to yield a non-negligible population of primordial SMBHs while satisfying bounds from CMB spectral distortions. While it is unclear whether primordial SMBHs can appreciably form in any viable single-field models, it is plausible that the necessary heavy tails can arise in certain multi-field scenarios. In particular, curvaton models have long been known to generate density perturbations with sizeable non-Gaussianities in the case that the curvaton remains subdominant at the time of its decay \cite{Sasaki:2006}. While the standard curvaton with quadratic potential is incapable of producing the dramatic non-Gaussianities needed for our scenario, there is good reason to believe that these can be augmented for a self-interacting curvaton model. We explore these possibilities in the following sections.

%

\section{Standard Curvaton Scenario}\label{standardcurvaton}

Curvaton models introduce a second light, unstable spectator field that is present during inflation and that is responsible for generating the dominant contribution to the primordial curvature perturbations~\cite{Lyth:2001,Moroi:2001,Enqvist:2001,Bartolo:2002}. The perturbations of the curvaton are initially isocurvature, but become adiabatic upon curvaton decay sometime after inflation ends~\cite{Mollerach:1990}. Due to the non-linearity inherent in this transfer, the full perturbation $\zeta$ can become quite non-Gaussian. In particular, when the curvaton is still very subdominant at decay, the inefficient conversion can yield a very heavy-tailed distribution for $\zeta$. 

Non-Gaussianity in the curvaton model was first investigated using the $\delta N$ formalism in Ref.~\cite{Sasaki:2006}, with implications for PBH formation analyzed in Refs.~\cite{Young:2013,Pi:2021,Gow:2023}. As we will see, the standard curvaton model with only a quadratic potential cannot produce sufficient non-Gaussianity to generate a non-negligible abundance of SMBHs without violating spectral distortions constraints.

\subsection{Curvaton Cosmology}

We begin by reviewing the calculation of the curvature perturbation $\zeta$ and its statistics in the standard curvaton scenario \cite{Sasaki:2006}, for which the total potential is
\begin{equation}
    V(\phi,\chi) = V_\phi(\phi) + V_\chi(\chi) \,,
\end{equation}
where $V_\phi$ is the unspecified potential of the inflaton $\phi$, and $V_\chi = m_\chi^2 \chi^2/2$ is the quadratic potential of the curvaton $\chi$. The curvaton mass is required to be light, satisfying $m_\chi \ll H$ throughout inflation, such that quantum perturbations are the dominant influence on its evolution. This also implies that the background field $\bar{\chi}$ will remain effectively fixed at some initial value $\bar{\chi}_*$ during inflation, where a star denotes the value of quantities at horizon exit. For the curvaton energy density to remain subdominant throughout inflation, we demand $\bar{\chi}_* \ll \sqrt{ 2 V_\phi/m_\chi^2}$. Just like the inflaton, the curvaton receives perturbations $\delta \chi_* \simeq H_*/2\pi$ set by the Hubble rate at horizon exit $H_*$. Since the curvaton is a weakly coupled field, we expect the perturbations $\delta \chi_*$ to be described by a Gaussian random field. Thus we can write the curvaton at horizon exit as the sum of a background field and a linear perturbation, with no higher order terms:
\begin{equation}
    \chi_* = \bar{\chi}_* + \delta \chi_* \,.
\end{equation}
The goal of this section will be to relate these initial Gaussian field perturbations to the total curvature perturbation $\zeta$ via some mapping $\zeta = f(\delta \chi_*)$. This will be the key to constructing the probability distribution function for $\zeta$, since the statistics of a non-Gaussian variable are completely determined by the statistics of a Gaussian reference variable when the mapping between them is specified. 

When inflation ends and the inflaton decays, the universe enters into an era of radiation domination, with $\rho_{R} \sim a^{-4}$. At this point, the curvaton energy density is subdominant and its fluctuations are still isocurvature in nature. As the Hubble rate decreases, it eventually drops below $m_{\chi}$, causing the curvaton to start oscillating about the minimum of its potential. We denote the field value at which this occurs by $\bar{\chi}_{0}$. During this oscillating phase, the curvaton redshifts like matter with $\rho_\chi \propto a^{-3}$ and its energy density grows linearly relative to radiation. Finally, when $H \sim \tau^{-1}$, where $\tau$ is the $\chi$ lifetime, the curvaton decays to radiation and its isocurvature perturbations become adiabatic perturbations, assuming the decay products thermalize with the existing radiation.

\subsection{The $\delta N$ Formalism}

To calculate the distribution of the curvature perturbations in this model,
we employ the $\delta N$ formalism~\cite{Salopek:1990,Starobinsky:1982,Sasaki:1995,Sasaki:1998,Lyth:2004,Lyth:2005,Langlois:2008}, which we review in Appendix~A. This technique identifies $\zeta$ on large scales ($k \ll aH$) with the variation of inflationary $e$-folds across Hubble patches and non-perturbatively captures its non-Gaussianities.
The $\delta N$ formalism was first used to study non-Gaussianity in curvaton models in Refs.~\cite{Bartolo:2003,Sasaki:2006}. On a general hypersurface of uniform curvaton density, the conserved curvaton curvature perturbation $\zeta_\chi$ is~\cite{Bartolo:2003,Lyth:2004}
\begin{equation}\label{zetachi}
    \zeta_\chi (t, \vec{x}) = \delta N(t,\vec{x}) + \frac{1}{3} \ln \left( \frac{\rho_\chi(t,\vec{x})}{\bar{\rho}_\chi(t)} \right) \,,
\end{equation}
where $\delta N(t, \vec{x})$ is the perturbed number of $e$-folds, $\rho_\chi(t,\vec{x})$ is the $\chi$ energy density, and $\bar{\rho}_\chi(t)$ is its background value. In spatially flat slicing, this becomes
\begin{equation}\label{spatiallyflat}
    \zeta_\chi (t, \vec{x}) = \frac{1}{3} \ln \left( \frac{\rho_\chi(t,\vec{x})}{\bar{\rho}_\chi(t)} \right) \,,
\end{equation}
and the curvaton energy density can be written as
\begin{equation}
\label{zeta_chi1}
    \rho_\chi(t,\vec{x}) = e^{3 \zeta_\chi(t,\vec{x})} \, \bar{\rho}_\chi(t) \,.
\end{equation}  
In uniform total density slicing, Eq.~(\ref{zetachi}) becomes
\begin{equation}
\label{zeta_chi2}
    \zeta_\chi (t, \vec{x}) = \zeta + \frac{1}{3} \ln \left( \frac{\rho_\chi^{(u)}(t,\vec{x})}{\bar{\rho}_\chi^{(u)}(t)} \right) \,,
\end{equation}
where $\zeta$ is the total curvature perturbation.\footnote{Note that this generically has non-vanishing mean, $\expval{\zeta} \neq 0$, and so when we later consider PBH formation, we will have to define a physical $\zeta_{\rm phys} \equiv \zeta - \expval{\zeta}$ \cite{Pi:2021}. The expectation value $\expval{\zeta}$ can be computed using the Gaussian $P_{\delta \chi}$ as $\expval{\zeta} = \int d \delta_\chi \, \zeta \, P_{\delta_\chi}$, with $\zeta$ expressed as a function of $\delta_\chi$ given in Eq.~(\ref{lnX}).} Since $\zeta$ and $\zeta_\chi$ are gauge invariant quantities,
 Eq.~(\ref{zeta_chi1}) can be equated with Eq.~(\ref{zeta_chi2}) to yield
\begin{equation}\label{rhochiuniform}
    \rho_\chi^{(u)}(t,\vec{x}) = e^{3( \zeta_\chi - \zeta)} \bar{\rho}_\chi^{(u)}(t) \,.
\end{equation}
A similar treatment for the radiation energy density in uniform total density slicing gives
\begin{equation}\label{rhoraduniform}
    \rho_{R}^{(u)}(t,\vec{x}) = e^{4( \zeta_{R} - \zeta)} \bar{\rho}_{R}^{(u)}(t) \simeq e^{-4 \zeta} \bar{\rho}_{R}^{(u)}(t) \,,
\end{equation}
where we have assumed for simplicity that the main contribution to the curvature perturbation comes from the curvaton. In order to derive analytic results, we work in the instantaneous decay approximation such that the curvaton decays when $H = \tau^{-1}$, where $\tau$ is the curvaton lifetime. On a uniform total density slice at $t = \tau$, the energy densities satisfy
\begin{equation}
    \rho_{R}^{(u)}(\tau, \vec{x}) + \rho_\chi^{(u)}(\tau, \vec{x}) = \bar{\rho}^{(u)}(\tau) \,,
\end{equation}
where $\bar{\rho}^{(u)} = \bar{\rho}^{(u)}_{R} + \bar{\rho}^{(u)}_\chi$ is the total homogeneous energy density. Substituting Eqs.~(\ref{rhochiuniform}) and~(\ref{rhoraduniform}), this becomes a 4$^{\rm th}$ degree algebraic equation for $\zeta$ at $\tau$:
\begin{equation}
    e^{4 \zeta} - \left( e^{3 \zeta_\chi} \, \Omega_{\chi,{\tau}} \right) e^{\zeta} + \left( \Omega_{\chi,{\tau}} - 1 \right) = 0 \,.
\end{equation}
Alternatively, it is customary to introduce the parameter $r_{\tau}$, defined as \cite{Lyth:2001}:
\begin{equation}
\label{rtau}
    r_{\tau} = \frac{3 \Omega_{\chi,{\tau}}}{4 - \Omega_{\chi,{\tau}}} = \frac{3 \bar{\rho}_\chi^{(u)}}{3 \bar{\rho}_\chi^{(u)} + 4 \bar{\rho}_{R}^{(u)}} \bigg|_{\tau} \,,
\end{equation}
in terms of which the equation for $\zeta$ becomes:
\begin{equation}\label{solveforroots}
    e^{4 \zeta} - \frac{4 r_{\tau}}{3+r_{\tau}} \left(e^{3 \zeta_\chi} \right) e^{\zeta} + \frac{3r_{\tau} -3}{3+r_{\tau}} = 0 \,.
\end{equation}
The general solution is
\begin{equation}\label{lnX}
    \zeta = \ln X \,, \,\,\, X = \frac{B^{1/2} + \sqrt{A r_{\tau} B^{-1/2} - B}}{(3+r_{\tau})^{1/3}} \,,
\end{equation}
where $A = e^{3 \zeta_\chi}$ and we have defined
\be
    B &=& \frac{1}{2} \left[ C^{1/3} + (r_{\tau}-1)(3+r_{\tau})^{1/3} C^{-1/3} \right]  \\[6pt] 
    C &=& (Ar_{\tau})^2 + \sqrt{(Ar_{\tau})^4 + (3+r_{\tau})(1-r_{\tau})^3} \,.
\ee
This gives the mapping $\zeta = \ln [X(\zeta_\chi)]$ between $\zeta$ and $\zeta_\chi$.

\subsection{Calculating the Probability Distribution}

\begin{figure}[t!]
\centering
\includegraphics[width=0.45\textwidth]{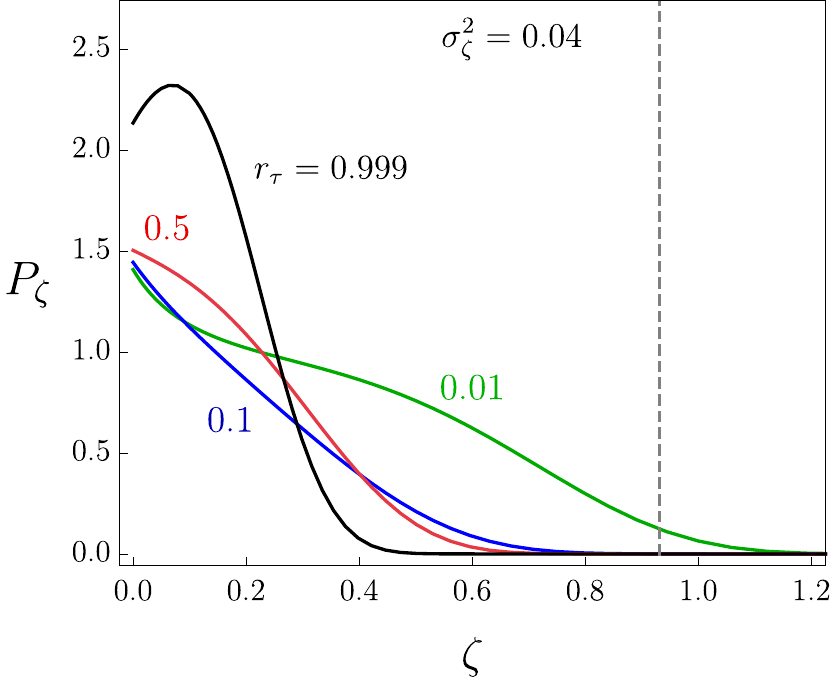}~~
\includegraphics[width=0.4655\textwidth]{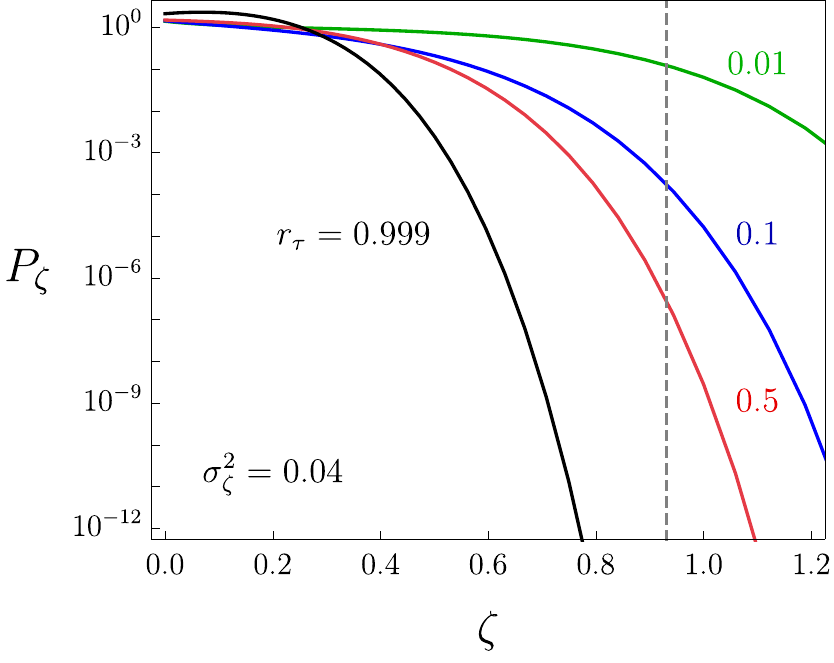}
\hspace{2mm}
\caption{The probability distribution function for the curvature perturbation $\zeta$ as given by Eq.~(\ref{zetapdf}) for various choices of $r_{\tau}$. The vertical dashed line corresponds to the threshold value for collapse $\zeta_c$, and the PBH mass fraction $\beta$ is obtained by integrating beyond this threshold. We see that a smaller $r_{\tau}$ corresponds to a heavier-tailed distribution, leading to a larger PBH abundance.
Note that the reference value $\sigma^2_\zeta = 0.04$ is chosen to illustrate the heaviness
of the tail as $r_\tau$ is varied, but the minimal scenario presented in Sec.~\ref{standardcurvaton}
cannot realize such a large value without violating the observational limits on 
the power spectrum shown in Fig.~\ref{SDconstraints}. }
\label{rdependence}
\end{figure}

Finally, to obtain $\zeta_\chi$ in terms of $\delta \chi_*$, we return to Eq.~(\ref{spatiallyflat}), which gave the curvature perturbation in spatially flat slicing as a function of perturbed and background energy densities. For the simple quadratic potential of this model, we have $\rho_\chi(t, \vec{x}) = m_\chi^2 \chi^2 /2$. Expanding $\chi(t, \vec{x}) = \bar{\chi}(t) + \delta \chi(t, \vec{x})$, we can write this as
\begin{equation}
    \rho_\chi(t, \vec{x}) = \frac{1}{2} m_\chi^2 \bar{\chi}^2 \left( 1 + \frac{\delta \chi}{\bar{\chi}} \right)^2 = \bar{\rho}_\chi(t) (1 + \delta_\chi)^2 \,,
\end{equation}
where $\delta_\chi = \delta \chi/\bar{\chi}$ is the curvaton contrast in spatially flat slicing. Comparing with $\rho_\chi = e^{3 \zeta_\chi} \bar{\rho}_\chi$, we see we should identify $e^{3 \zeta_\chi} = (1 + \delta_\chi)^2$. Finally in order to relate the contrast $\delta_\chi$ to its value at horizon exit, consider the equations of motion for $\chi$ and the perturbation $\delta \chi$:
\begin{subequations}
\label{EOM-chi}
\begin{equation}
    \frac{d^2 \bar{\chi} }{dt^2}  + 3 H   \frac{d \bar{\chi} }{dt}  + m_\chi^2 \bar{\chi} = 0 \,,
\end{equation}
\begin{equation}
       \frac{d^2  }{dt^2} (\delta \chi)  + 3 H \frac{d }{dt} (\delta \chi) + \left( \frac{k^2}{a^2} + m_\chi^2 \right) \delta \chi = 0 \,.
\end{equation}
\end{subequations}
On superhorizon scales $k \ll a H$ is negligible and so these reduce to the same equation. This implies $\delta \chi \sim \bar{\chi}$, such that $\delta \chi/\bar{\chi} = \delta \chi_*/\bar{\chi}_*$ and
\begin{equation}\label{Adef}
    A = e^{3 \zeta_\chi} = (1 + \delta_\chi)^2 = \left( 1 + \frac{\delta \chi_*}{\bar{\chi}_*} \right)^2 \,.
\end{equation}
Combining with Eq.~(\ref{lnX}), we obtain the desired mapping between $\zeta$ and the Gaussian initial field perturbations $\delta \chi_*$. 
%
%
%
We can now use probability conservation to write
\begin{equation}
\label{zetapdf}
    P_\zeta[\zeta] = P_{\delta_\chi} \left[\delta_\chi^{+}(\zeta) \right] \bigg| \frac{d \delta_\chi^{+}}{d\zeta} \bigg| + P_{\delta_\chi} \left[\delta_\chi^{-}(\zeta) \right] \bigg| \frac{d \delta_\chi^{-}}{d\zeta} \bigg| \,, 
\end{equation}
where $P_{\delta_\chi}$ is fully determined by the Gaussian variance $\sigma_0^2$ and the roots $\delta^{\pm}_\chi(\zeta)$ satisfy 
\begin{equation}\label{roots}
    \delta_\chi^{\pm} = -1 \pm \sqrt{    \left(    \frac{3+r_{\tau}}{4r_{\tau}}  \right) e^{3 \zeta} + \left(
    \frac{3r_{\tau}-3}{4r_{\tau}}  \right) e^{- \zeta}} \,,
\end{equation}
which arise from solving Eq.~(\ref{solveforroots}) and substituting Eq.~(\ref{Adef}). In Fig.~(\ref{rdependence}), we plot the probability distribution for the curvature perturbation as given by Eq.~(\ref{zetapdf}) for a few choices of $r_\tau$, which controls the heaviness of the tail.

\section{Heavy Tails and Primordial Black Holes in Curvaton Models}\label{nonstandard}

Viably forming an appreciable number of primordial SMBHs requires both amplified power on small scales and a  departure from Gaussianity. Thus we require the following two additional ingredients:
\begin{itemize}
\item{\bf Enhanced Power:} 
One mechanism for enhancing the power spectrum is to introduce a non-canonical kinetic term for the curvaton, which depends on the inflaton's field value \cite{Pi:2021}. In Sec.~\ref{curvatonPBH}, we review this scenario and calculate the power spectrum resulting from such a kinetic term. 

\item{\bf Large Non-Gaussianity:}
The non-Gaussianity in a curvaton model can be amplified through self-interactions, which lead to non-linear growth of $\chi$ perturbations between horizon exit and the onset of curvaton oscillations~\cite{Enqvist:2008,Enqvist:2009,Enqvist:2009sd,Taanila:2010,Enqvist:2010,Byrnes:2010,Fonseca:2011,Byrnes:2011,Kobayashi:2012}.
In Sec.~\ref{interactingcurvaton}, we review the evolution of the non-quadratic curvaton and calculate the resulting probability distribution function.

\end{itemize}

\subsection{Non-Minimal Curvaton Scenario}\label{curvatonPBH}

The minimal curvaton model described in Sec. \ref{standardcurvaton}  provides non-Gaussian statistics, but does not amplify $\mathcal{P}_\zeta$. This deficiency can be remedied with a non-canonical kinetic term\footnote{Such a term naturally arises in many dilatonic and axionic models of inflation~\cite{Domenech:2015}.} for the curvaton~\cite{Pi:2021}:
\begin{equation}
    \mathcal{L} \supset \frac{1}{2} f(\phi)^2 (\partial \chi)^2 \,.
\end{equation}
If $f(\phi)$ is chosen such that this kinetic term is suppressed at field values $\phi = \phi_*$, the power spectrum will be enhanced on scales corresponding to the horizon size at $\phi_*$. 
For concreteness, consider the evolution of the inflaton and curvaton governed by the following system of equations:
\be
    \ddot{\phi} + 3 H \dot{\phi} + V'(\phi) = f f' \dot{\bar{\chi}}^2 \simeq 0 \,,~~
    \ddot{\bar{\chi}} + \left( 3 H + \frac{2 f'}{f} \dot{\phi} \right) \dot{\bar{\chi}} + \frac{m_\chi^2}{f^2} \bar{\chi} = 0 \,,
\ee
where $f' \equiv \partial_\phi f$ and the source term of the first equation is negligible until the curvaton begins to oscillate.
Similarly, the curvaton perturbation evolves according to 
\be
\frac{d^2}{dt^2}(\delta \chi) + \left( 3 H +  \frac{2 f'}{f} \dot{\phi} \right) \frac{d}{dt}(\delta \chi)  + \left( \frac{k^2}{a^2} + \frac{m_\chi^2}{f^2} \right) \delta \chi \simeq 0 \, ,
\ee
 which, to leading order, can be simplified inside the horizon, $k \gg aH$, to yield 
\begin{equation}
\label{sol-inside}
    \frac{d^2}{dt^2} \left( f  \delta \chi \right) + \frac{k^2}{a^2} (f \delta \chi) \simeq 0 \,,
\end{equation}
 whose solution can be written as 
\begin{equation}
    \delta \chi \simeq \frac{1}{\sqrt{2 k} \, a f} \exp\left(-i k \! \int \frac{dt}{a}  \right) ,
\end{equation}
which establishes the initial conditions in the adiabatic vacuum. As in Sec.~\ref{standardcurvaton}, on superhorizon scales, $k \ll a H$,
$\bar{\chi}$ and $\delta \chi$ evolve according to the same equation, so their solutions have the same functional form for all $t > t_*$:
\begin{equation}
\label{contrast}
    \frac{\delta \chi}{\bar{\chi}} = \frac{\delta \chi_*}{\bar{\chi}_*} \simeq \frac{H_*}{\sqrt{2k^3} f(\phi_*) \bar{\chi}_*} \,,
\end{equation}
where $k = a(t_*) H(t_\star)$. Using Eq.~(\ref{contrast}), the power spectrum for $\delta_\chi$ becomes
\begin{equation}
\label{finalspectrum}
    \mathcal{P}_{\delta_\chi}(k) = \frac{k^3}{2 \pi^2} \left| \frac{\delta \chi_k}{\bar{\chi}} \right|^2 = \frac{1}{\bar{\chi}_*^2} \left( \frac{H_*}{2 \pi f(\phi_*)} \right)^2 ~, 
\end{equation}
so, if $f(\phi)$ is chosen to have a dip at $\phi_*$, corresponding to $k_* = k_{\rm BH}$, $\mathcal{P}_{\delta_\chi}$ will
exhibit a peak at $k_{\rm BH}$. However, for modes far away from $k_{\rm BH}$, $f(\phi_*) \approx 1$, recovering the nearly scale-invariant spectrum required for consistency with CMB observations on larger scales. Assuming $f(\phi)$ has such a localized feature, 
combining Eqs.~(\ref{zetapdf}), (\ref{finalspectrum}), and (\ref{betaestimate}), the PBH abundance becomes
\be
    \beta  = 2 \int_{\delta_{\chi, {c}}^{+}}^\infty d \delta_\chi \, P_{\delta_\chi}[\delta_\chi] + 2 \int_{-\infty}^{\delta_{\chi, {c}}^{-}} d \delta_\chi \, P_{\delta_\chi} [\delta_\chi] 
    = \text{erfc} \left( \frac{\delta_{\chi, {c}}^{+}}{\sqrt{2} \sigma_0} \right) + \text{erfc} \left( \frac{| \delta_{\chi, {c}}^{-} |}{\sqrt{2} \sigma_0} \right) ~,
\ee
where $\delta_{\chi, {c}}^{\pm} = \delta_{\chi}^{\pm}(\zeta_c)$ are the roots of Eq. (\ref{roots}) evaluated at the threshold.

Using Eq.~(\ref{SDestimates}), we obtain the maximum value of $\beta$ consistent with spectral distortion constraints.\footnote{Note that the variance $\sigma_\zeta^2$ corresponds to the physical curvature perturbation, $\sigma_\zeta^2 = \langle \zeta_{\rm phys}^2 \rangle = \expval{\zeta^2} - \expval{\zeta}^2$, which can be computed from $\zeta(\delta_\chi)$ in Eq.~(\ref{lnX}) as $\sigma_\zeta^2 = \int d \delta_\chi \, \zeta^2 \, P_{\delta_\chi} - \left( \int d \delta_\chi \, \zeta \, P_{\delta_\chi} \right)^2$.}
\begin{figure}[t!]
\centering
\includegraphics[width=0.77\textwidth]{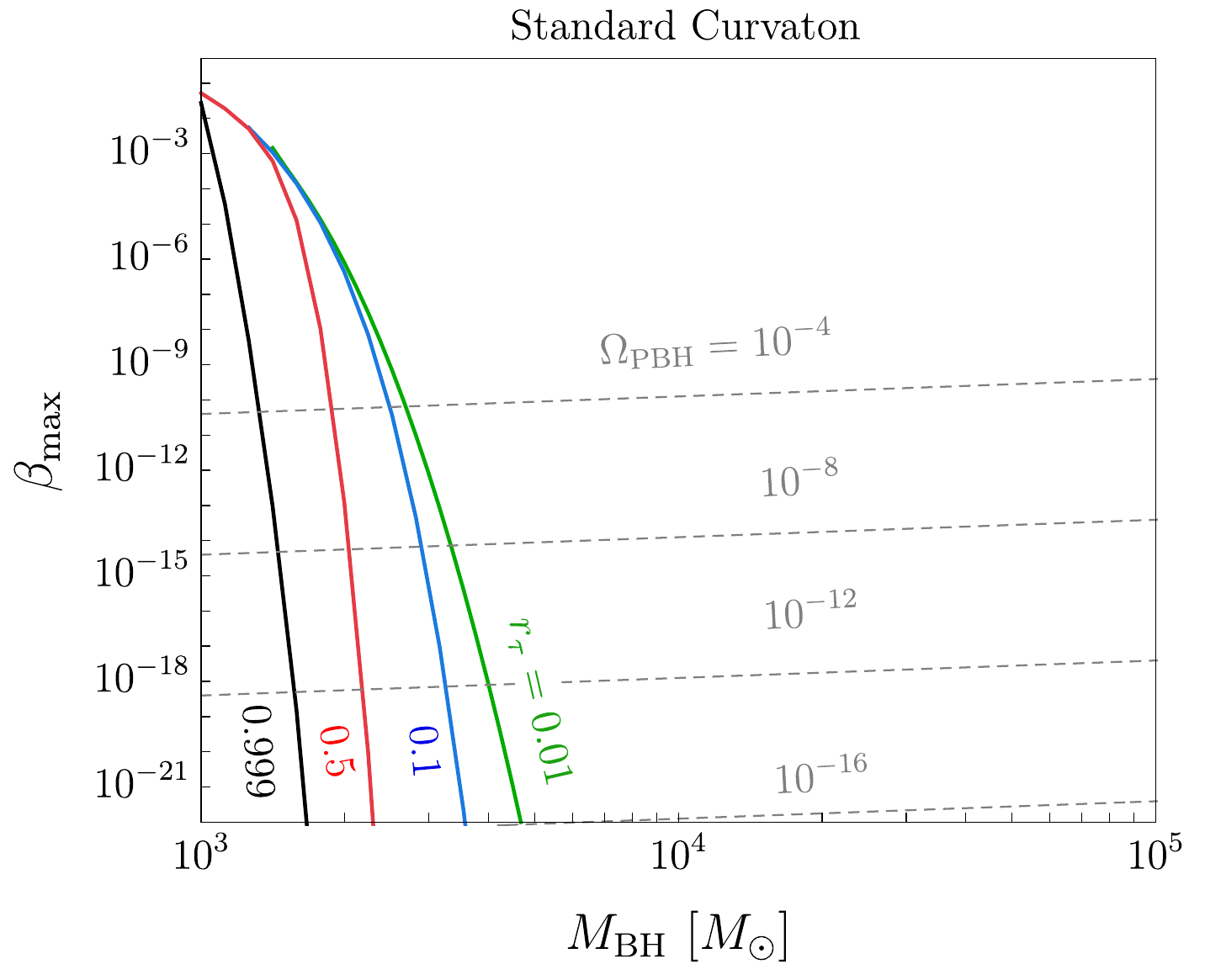}
\hspace{-1cm}
\caption{The maximal PBH mass fraction at formation $\beta_{\rm max}$ in the standard curvaton scenario for $\sigma_\zeta^2$ subject to spectral distortion constraints for various choices of $r_{\tau}$ from Eq.~(\ref{rtau})}
\label{standardcurvbeta}
\end{figure}
The degree of non-Gaussianity in this scenario is governed by $r_{\tau}$, as defined in Eq.~(\ref{rtau}). In the $r_{\tau} \rightarrow 1$ limit, the curvaton dominates the energy density prior to its decay, so the relation between $\zeta$ and $\zeta_{\chi}$ from Eq.~(\ref{solveforroots}) is approximately linear. Thus, in this regime, the non-Gaussianity comes entirely from the non-linear relationship between $\zeta_\chi$ and $\delta_\chi$; see Eq.~(\ref{Adef}). In the opposite regime\footnote{It may seem counterintuitive that a very subdominant curvaton can still generate the curvature perturbation. The key is that because the curvaton is subdominant during inflation, its perturbations from quantum fluctuations are large relative to the background field value. Ref.~\cite{Enqvist:2009zf} finds that a subdominant curvaton can still viably produce the curvature perturbation provided $r_\tau \gtrsim 10^{-3}$.} that the curvaton is still very subdominant when it decays ($r_{\tau} \ll 1$), the relation between $\zeta$ and $\zeta_\chi$ is highly non-linear, and so the degree of non-Gaussianity is large. This is reflected in Figs.~\ref{rdependence} and \ref{standardcurvbeta}. Clearly this scenario is incapable of producing primordial SMBHs while satisfying spectral distortion constraints.

\subsection{Self-Interacting Curvaton Scenario}\label{interactingcurvaton}

Introducing curvaton self-interactions allows for non-linear evolution of $\delta_\chi$ between horizon exit and the onset of oscillations, which can lead to even more dramatic departures from Gaussianity. This scenario is also physically well-motivated since the curvaton needs to decay for the isocurvature perturbations to be converted to adiabatic perturbations. The effects of self-interactions in curvaton models were first investigated in Refs.~\cite{Enqvist:2008,Enqvist:2009,Enqvist:2009sd,Taanila:2010,Enqvist:2010,Byrnes:2010,Fonseca:2011,Byrnes:2011,Kobayashi:2012}. However, these studies restricted themselves to computing the non-linearity parameters $f_{\rm NL}$ and $g_{\rm NL}$, which do not capture the non-perturbative statistics in the tail of the distribution. Unfortunately following the curvaton's evolution in the non-quadratic regime and computing the exact resulting curvature distribution require extensive numerics beyond the scope of this study. We offer here a schematic picture, but leave the detailed model building to future studies.



We now allow the curvaton potential $V(\chi)$ to be an arbitrary well-defined function of $\chi$. The cosmological evolution of the curvaton proceeds similarly to the case of the quadratic potential, but with a few crucial differences. At a time $t = t_{\rm int}$ corresponding to  $V'(\chi_*) \sim H$, $\chi$ begins rolling towards the minimum of its potential. We choose parameters such that this occurs while the curvaton's energy density is dominated by the interaction terms. At a later time  $t = t_0 > t_{\rm int}$, these interaction terms become subdominant and the curvaton mass term drives field evolution, resulting in matter-like scaling $\rho_\chi \propto a^{-3}$. Note that with self-interactions, the curvaton energy density generically falls off faster than in the quadratic case, resulting in a smaller $r_\tau$ at the time of decay. 

Recall that in the case of the quadratic potential, the curvaton density contrast $\delta_\chi$ remained constant after horizon exit since the background field $\bar{\chi}$ and perturbation $\delta \chi$ obeyed the same equation of motion, shown in Eq.~(\ref{EOM-chi}). This led to $\delta \chi/\bar{\chi} = \delta \chi_*/\bar{\chi}_*$, which allowed $\delta_\chi$ to be used as a Gaussian reference variable. Upon introducing interactions this is no longer the case, as $\delta_\chi$ evolves non-trivially between $t_{\rm int}$ and the onset of quadratic oscillations at $ t_0$. In this regime, the equation of motion for the perturbation is:
\be
\label{deltachieq}
    \frac{d^2}{dt^2} (\delta \chi) + 3 H \frac{d}{dt} (\delta \chi) + V''_\chi \delta \chi = 0 \,,
\ee
where it is understood that the second derivative of the potential should be evaluated at the background field value. 

A natural choice of Gaussian reference variable is the initial curvaton perturbation $\delta \chi_*$, which can be related to the $\zeta_\chi$ by first solving Eq.~(\ref{deltachieq}) along with the equation for the background field $\bar{\chi}$, computing the total and background energy densities at decay, and finally applying Eq.~(\ref{spatiallyflat}). The resulting $\delta \chi_*^i = g_i(\zeta_\chi)$ can then be mapped onto the total curvature perturbation $\zeta$ via Eq.~(\ref{solveforroots}), since the relationship between $\zeta_\chi$ and $\zeta$ is unchanged in the presence of $\chi$ self-interactions. Although
an exact solution requires the use of numerical techniques, an approximate relation can be derived in the limit of weak interactions. 

We are interested in the relation between $\delta \chi_*$ and $\zeta_\chi$ at the time of curvaton decay. For $\chi$ values sufficiently close to the minimum of its potential,   the potential is approximately quadratic, and the energy density is
\begin{equation}
    \rho_\chi \simeq \frac{1}{2} m_\chi^2 \chi_{0}^2 \,,
\end{equation}
where $\chi_{0}$ is the amplitude at the onset of oscillations. Since non-linear evolution takes place between horizon exit and oscillation, this initial amplitude is a function of initial field value $\chi_*)$. In
terms of background field values $\bar{\chi}_{0} \equiv \chi_{0}(\bar{\chi}_*)$, this can be expanded as:
\begin{equation}
    \chi_{0} = \bar{\chi}_{0} + \sum_{n=1} \frac{1}{n!} \bar{\chi}_{0}^{(n)} \delta \chi_*^n ~~,~~
    \bar{\chi}_{0}^{(n)} \equiv \frac{\partial^n \chi_{0}}{\partial \chi_*^n} \Big|_{\chi=\bar{\chi}_*},
\end{equation}
and the energy density can then be written as:
\begin{equation}
    \rho_\chi \simeq \bar{\rho}_\chi \left[ 1 + \frac{1}{\bar{\chi}_{0}} \sum_{n=1} \frac{1}{n!} \bar{\chi}_{0}^{(n)} \delta \chi_*^n \right]^2 \,,
\end{equation}
where $\bar{\rho}_\chi = \frac{1}{2} m_\chi^2 \bar{\chi}_{0}^2$. Comparing with Eq. (\ref{spatiallyflat}), the bracketed quantity is identified with $e^{3 \zeta_\chi}$. Then also expanding
\be
\zeta_{\chi} = \zeta_{\chi,1} + \sum^\infty_{n=2}\frac{1}{n!} \zeta_{\chi,n}~~~,
\ee
we can write to leading order:
\begin{equation}\label{intzetachi}
\begin{split}
    \zeta_\chi(\delta \chi_*) & = \frac{2}{3} \left(  \frac{\bar{\chi}'_{0}}{\bar{\chi}_{0}} \right) \delta \chi_* + \frac{1}{3} \left( \frac{\bar{\chi}_{0} \bar{\chi}''_{0}}{{\bar{\chi}}^{\prime\,2}_{0}} - 1 \right) \left( \frac{\bar{\chi}'_{0}}{\bar{\chi}_{0}} \right)^2 \delta \chi_*^2\\
    & \hspace{9.5mm} + \frac{1}{9} \left( \frac{\bar{\chi}_{0}^2 \bar{\chi}'''_{0}}{{\bar{\chi}}^{\prime\,3}_{0}} - 3 \frac{\bar{\chi}_{0} \bar{\chi}''_{0}}{{\bar{\chi}}^{\prime\,2}_{0}} + 2 \right) \left( \frac{\bar{\chi}'_{0}}{\bar{\chi}_{0}} \right)^3 \delta \chi_*^3 \,,
\end{split}
\end{equation}
which can then be substituted into Eq. (\ref{lnX}) to obtain $\zeta$ as a function of the Gaussian reference variable $\delta \chi_*$. Note that though we have written this expression to fixed order, higher order terms can become significant in the presence of large non-Gaussianities. It is also possible to invert this mapping to obtain the roots $g_i(\zeta) = \delta \chi_*^i(\zeta)$. The probability distribution is then:
\begin{equation}
    P_\zeta[\zeta] = \sum_{j} \left| \frac{g_j(\zeta)}{d\zeta} \right| P_{\delta \chi_*}[g_j(\zeta)] \,,
\end{equation}
where the sum runs over all real roots. 

\begin{figure}[t!]
\centering
\includegraphics[width=0.77\textwidth]{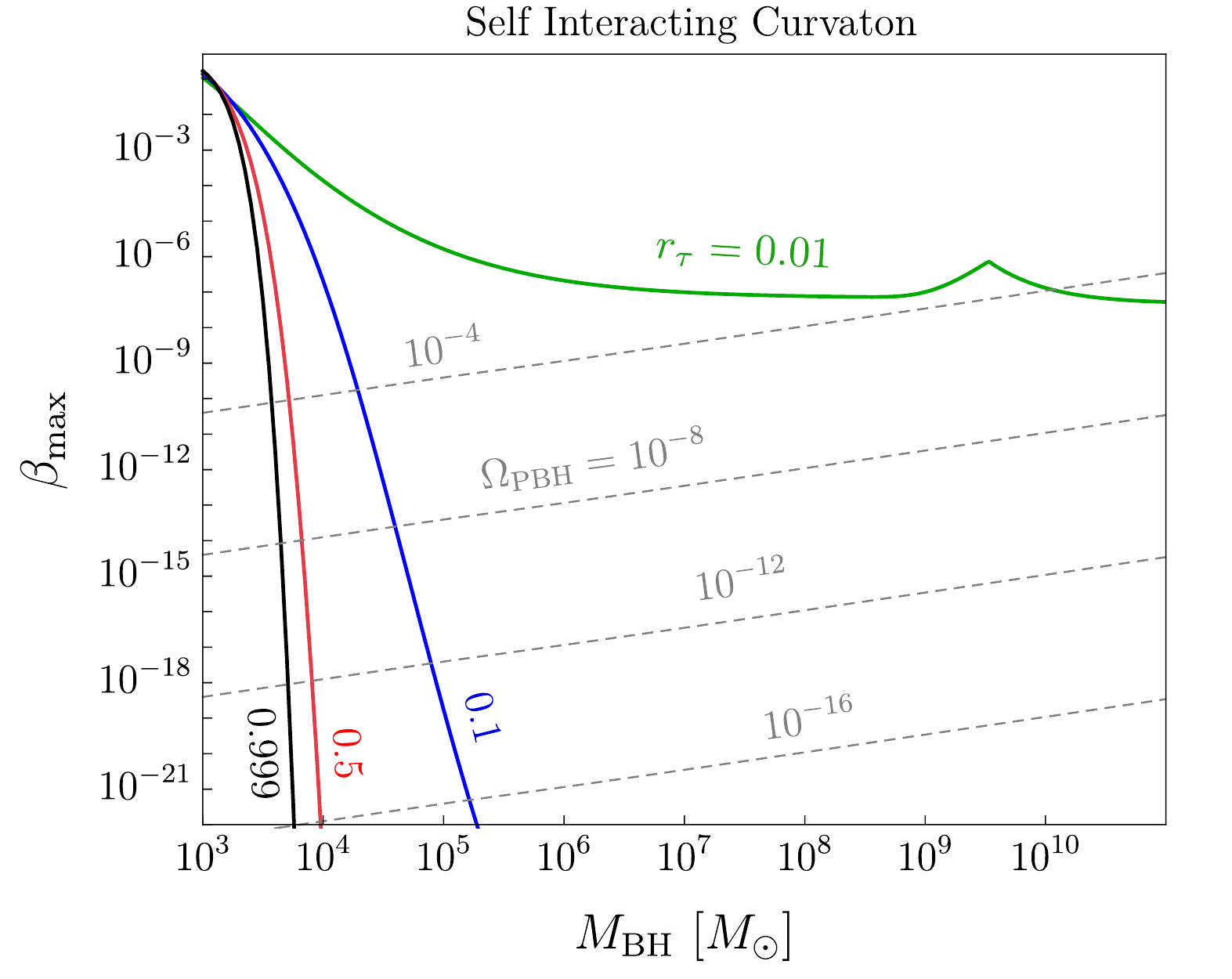} 
\hspace{1cm}
\caption{Maximal PBH mass fraction at formation $\beta_{\rm max}$ as a function of PBH mass in the self-interacting curvaton model from Sec. \ref{interactingcurvaton}, where $\sigma_\zeta^2$ saturates spectral distortion bounds for different values of $r_{\tau}$, as defined in Eq.~(\ref{rtau}). We take $\zeta_\chi = c_1\, \delta \chi_* + c_2 \, \delta \chi_*^2 + c_3 \, \delta \chi_*^3$ to take the form of Eq.~(\ref{intzetachi}), with sample parameters fixed as $(c_1,c_2,c_3) = (1, 0.5, 0.1)$. Contours of constant $\Omega_{\rm BH}$ today are shown in dashed gray.}
\label{betaplot}
\end{figure}

In Fig.~\ref{betaplot}, we plot the maximal PBH abundance consistent with spectral distortion constraints for a sample set of parameters. Comparison with $\beta_{\rm max}$ in the standard curvaton scenario (see Fig.~\ref{standardcurvbeta}) reveals the dramatic amplification of non-Gaussianity that self-interactions can afford. We leave the determination of a potential that could actually yield these parameter values to future investigation.
%
%

\section{Discussion and Conclusions}\label{conclusions}

Much remains to be understood about the origin and evolution of our universe's most massive black holes. The inferred population of supermassive black holes with $M_{\rm BH} \sim 10^6 - 10^{11} M_\odot$ in the high-redshift universe is perhaps surprising, and challenges the standard assumption that these objects formed from low mass seeds which grew through the processes of accretion and mergers. In this study, we have taken this as motivation to consider the possibility that some of our universe's supermassive black holes may be primordial in origin, having formed from the direct collapse of overdensities seeded by inflation. 

Forming primordial SMBH from direct collapse requires an enhanced power spectrum on small scales ($k \sim {\rm 10-10^4 \, Mpc^{-1}}$), which results in dangerous CMB spectral distortions. Since the CMB exhibits a nearly perfect blackbody spectrum, such distortions exclude the possibility that a population of primordial SMBH could have formed from Gaussian density perturbations. However, if the distribution of primordial curvature perturbations were highly non-Gaussian, it is possible that primordial black holes may have formed from smaller peaks in the power spectrum.
%
%
%
To evade limits from spectral distortions, the tail of the probability distribution must be very  heavy, falling off as $P_\zeta \sim \exp\left(-|\zeta|^n \right)$ with $n \lesssim 0.6$; we are not aware of any single-field inflationary model that can realize this behavior. 

In this paper, we have explored the possibility of generating curvature perturbations with heavy-tailed probability distributions in the event that a self-interacting curvaton field is also present as a spectator during inflation. In the standard curvaton scenario, non-Gaussianity arises from the inefficient conversion of isocurvature perturbations into adiabatic perturbations when the curvaton decays. However, the degree of non-Gaussianity in this minimal realization is insufficient to yield an appreciable primordial SMBH population. Introducing curvaton self-interactions results in non-linear evolution of the curvaton contrast $\delta_\chi$ between horizon exit and the onset of quadratic oscillations, potentially resulting in a heavier-tailed distribution, as seen in Fig.~\ref{betaplot}. We leave the detailed model building and determination of a potential that can realize this benchmark point to future investigation. Provided such a potential can be found, the non-minimal self-interacting curvaton could viably yield SMBHs from the direct collapse of primordial perturbations, without violating spectral distortion constraints.

We note that beyond the bounds from CMB spectral distortions, there are two further more speculative constraints that potentially need to be addressed in a full model. First, as pointed out by Ref.~\cite{Ando:2022}, the small scale curvature perturbation impacts structure evolution as traced by dark matter halos. Presuming a bump in the small scale power spectrum, they investigated host halo and subhalo evolution. By comparing the results of their simulations with the observed number of dwarf spheroidal galaxies and stellar streams, they were able to place bounds on $\mathcal{P}_\zeta$ to greater sensitivity than COBE/FIRAS for a region of wavenumbers. The result would exclude some, but not all, of the SMBH mass range we consider. These bounds are less robust than those coming from CMB spectral distortion and are subject to modeling uncertainties, but nevertheless should be noted. Second, the large non-Gaussianities described in this model would likely lead to heavy clustering of quasars. Ref.~\cite{Shinohara:2023} calculated the observed angular correlation function using the 92 quasars at $z\sim 6$ reported by the Subaru High-$z$ Exploration of Low-Luminosity Quasars project and compared this with that predicted in a PBH model which used a light spectator field to source non-Gaussian primordial perturbations. They found the amplitude of the angular correlation function predicted by the model to be much larger than that observed, a result which would seem to preclude PBHs as the sole progenitors of SMBHs in this redshift range. More concretely, they were able to restrict the fraction of PBH originating SMBHs to $\lesssim 10^{-4}$.

These potential constraints notwithstanding, this work is especially timely in light of two distinct recent observations. The James Webb Space Telescope (JWST) has reported a number of surprisingly luminous high-redshift galaxy candidates \cite{Naidu:2022,Adams:2023,Labbe:2022,Finkelstein:2023} whose existence poses a challenge to the standard $\Lambda$CDM paradigm. These massive early galaxies could conceivably be explained if primordial black holes accelerated galaxy formation in the early universe \cite{Hutsi:2022}. Meanwhile, the NANOGrav collaboration and several other pulsar timing array experiments have just announced evidence of a signal consistent with the stochastic gravitational wave background in the nHz frequency range~\cite{NANOGrav:2023c,EPTA:2023,PPTA:2023,CPTA:2023}. The leading astrophysical interpretation of this signal is that it consists of gravitational waves from supermassive black hole binary mergers. However, some aspects of this data, such as the frequency scaling of the spectral density parameter, are not particularly well-fit by this interpretation~\cite{NANOGrav:2023a,NANOGrav:2023b}. Given that the distribution of SMBH binaries would be different if these objects were of a primordial origin, one avenue for future investigation would be to compute the gravitational wave signal predicted in this scenario. While for homogenously distributed primordial SMBHs this possibility is in conflict with constraints on total abundance from large scale structure and the UV galaxy luminosity function \cite{Gouttenoire:2023}, a highly clustered population could still viably source the gravitational wave signal \cite{Depta:2023}. Given that a clustered population can arise in the presence of non-Gaussianities, it would be interesting to see what our model predicts in this context; see Refs.~\cite{Shinohara:2021,Shinohara:2023} for closely related work. There are also other signals, such as scalar-induced secondary gravitational waves, which could offer complimentary evidence of this scenario and deserve further study; see Refs.~\cite{Franciolini:2023,Wang:2023} for a potential connection between scalar-induced gravitational waves and the PTA signal. Regardless, these recent observations provide us with strong motivation to better understand the cosmic origin of our universe's supermassive black holes.


\acknowledgments

We would like to thank Wayne Hu and Keisuke Inomata for helpful conversations. AI thanks the organizers and participants of the New Horizons in Primordial Black Hole Physics (NEHOP) workshop, where part of this work was completed. DH, GK and AS are supported by the Fermi Research Alliance, LLC under Contract No.~DE-AC02-07CH11359 with the U.S. Department of Energy, Office of Science, Office of High Energy Physics. 

\bibliographystyle{JHEP.bst}
\bibliography{references}


\appendix
\setcounter{equation}{0}
\setcounter{figure}{0}

\section{The $\delta N$ Formalism}\label{numericaldeltaN}

\subsection{Review}

The $\delta N$ formalism~\cite{Salopek:1990,Starobinsky:1982,Sasaki:1995,Sasaki:1998,Lyth:2004,Lyth:2005,Langlois:2008} is a technique for computing the non-linear curvature perturbation $\zeta$ on large scales by identifying it with the perturbed logarithmic expansion from some initial state to a final state of fixed energy density. In a homogenous background, the number of $e$-folds elapsed between two moments of times $t_1$ and $t_2$ is
\begin{equation}
    \bar{N}(t_2,t_1) = \int_{t_1}^{t_2}dt \, H \,.
\end{equation}
Meanwhile, the amount of expansion in a perturbed universe is \cite{Lyth:2004}
\begin{equation}
    N(t_2,t_1; \vec{x}) = \int_{t_1}^{t_2}dt \, (H + \dot{\psi}) \,,
\end{equation}
where $\psi$ is the curvature perturbation appearing in the decomposition of the spatial 3-metric $g_{ij} = a^2(t) e^{2\psi(t,\vec{x})} \delta_{ij}$.\footnote{We ignore tensor perturbations.} Note that this expression holds on superhorizon scales, where spatial gradients can be neglected. We define $\delta N \equiv N(t_2,t_1;x) - \bar{N}(t_2,t_1)$ to be the difference between the perturbed and unperturbed expansion, and see that it equates to the change in the curvature perturbation from an initial hypersurface at $t_1$ to a final hypersurface at $t_2$
\begin{equation}\label{deltaNNN}
    \delta N \equiv N(t_2,t_1; \vec{x}) - \bar{N}(t_2,t_1) = \psi(t_2, \vec{x}) - \psi(t_1, \vec{x}) \,.
\end{equation}
Equivalently, the change in $\psi$ going from one choice of slicing to another is the difference between the actual number of $e$-folds $N$ and the homogenous background value $\bar{N}$.

Since the curvature perturbation $\psi$ is a gauge-dependent quantity, whose value depends on the choice of slicing, it is convenient to combine this with the gauge-dependent density perturbation $\delta \rho$ to form the gauge-invariant curvature perturbation \cite{Bardeen:1980}:
\begin{equation}\label{generalzeta}
    \zeta(t,\vec{x}) = \psi(t,\vec{x}) + \frac{1}{3} \int_{\bar{\rho}(t)}^{\rho(t,\vec{x})} \frac{d\rho}{(1+w) \rho} \,,
\end{equation}
where $w$ is the equation of state of the cosmological fluid, $\rho(t,\vec{x})$ is the inhomogenous local energy density, and $\bar{\rho}(t)$ is the homogenous background energy density. It was demonstrated in Ref.~\cite{Lyth:2004} that this quantity is conserved on superhorizon scales.

We would like to equate this gauge-invariant, conserved curvature perturbation $\zeta$ with the perturbed logarithmic expansion $\delta N$. By choosing the initial hypersurface at $t_1$ to be spatially flat, such that $\psi(t_1, \vec{x}) = 0$, from Eq.~(\ref{deltaNNN}) we can equate $\delta N(t_2, t_1; \vec{x} = \psi(t_2, \vec{x})$. By choosing the finial hypersurface at $t_2$ to be uniform density, such that $\rho(t_2, \vec{x}) = \bar{\rho}(t_2)$, from Eq.~(\ref{generalzeta}) we have $\zeta(t_2,\vec{x}) = \psi(t_2, \vec{x})$. Combining these expressions gives the $\delta N$ formula
\begin{equation}
    \zeta(t_2,\vec{x}) = \delta N(t_2,t_1; \vec{x}) \,,
\end{equation}
relating the curvature perturbation with the perturbed expansion between a spatially flat and uniform energy density hypersurface.

\subsection{Numerical Implementation}

It is also possible to implement the $\delta N$ formalism numerically. Following inflaton decay, the exact system of equations describing the evolution of the curvaton and radiation bath is
\begin{equation}
    \ddot{\chi} + \left(3 H + \frac{1}{\tau} \right) \dot{\chi} + V_\chi^\prime = 0 ~~~,~~~
    \dot{\rho}_{R} + 4 H \rho_{R} = \frac{ \dot{\chi}^2 }{\tau} ~~~,~~~
    H^2 = \frac{8 \pi}{3 M_{\rm Pl}^2} (\rho_{R} + \rho_\chi) ~.~~~
\end{equation}
We set the initial conditions at the end of inflation by specifying $\chi_i = \chi_*$, $\dot{\chi}_i=0$, and $H_i = H_*$, which in turn determines $\rho_{R,i} = \frac{3 M_{\rm Pl}^2}{8\pi} H_i^2 - V_\chi(\chi_i)$. The system of equations should be evolved until a final time $t_f$ satisfying $H_f \ll \tau^{-1}$, such that the curvaton has decayed completely. The number of $e$-folds elapsed is then computed as $N = \ln (a_f/a_i)$. This procedure should then be repeated for the perturbed field value $\chi_* + \delta \chi_*$, with the fluctuation determined by the size of Hubble at horizon exit $\delta \chi_* = H_*/2\pi$. Evolving until the same final hypersurface of fixed energy density, the curvature perturbation is computed as:
\begin{equation}
    \zeta = N(\chi_* + \delta \chi_*) - N(\chi_*) \,.
\end{equation}
Repeating for many different $\delta \chi_*$ gives a functional relation between $\zeta$ and the Gaussian $\delta \chi_*$. We leave an in-depth numerical study to future investigation.

\end{document}